# Semiclassical theory of frequency combs generated by parametric modulation of optical microresonators

M. Sumetsky
Aston Institute of Photonics Technologies, Aston University, Birmingham B4 7ET, UK
Email: *m.sumetsky@aston.ac.uk*

**Abstract**
An optical microresonator, which parameters are periodically modulated in time, can generate optical frequency comb (OFC) spectral resonances equally spaced by the modulation frequency. Significant recent progress in realization of OFC generators based on the modulation of microresonator parameters boosted interest to their further experimental development and theoretical understanding of underlying phenomena. However, most of theoretical approaches developed to date were based on the lumped parameter models which unable to evaluate, analyse, and optimize the effect of spatial distribution of modulation inside microresonators. Here we develop the multi-quantum semiclassical theory of parametrically excited OFCs which solves these problems. As an application, we compare OFCs which are resonantly or adiabatically excited in a racetrack microresonator (RTM) and a SNAP (Surface Nanoscale Axial Photonics) bottle microresonator (SBM). The principal difference between these two types of microresonators consists in much slower propagation speed of whispering gallery modes along the SBM axis compared to the speed of modes propagating along the RTM waveguide axis. We show that, due to this difference, similar OFCs can be generated by an SBM with a much smaller size compared to that of the RTM. Based on the developed theory, we analytically express the OFC spectrum of microresonators through the spatial distribution of modulated parameters and optimize this distribution to arrive at the strongest OFCs generated with minimum power consumption.

## 1. Introduction

An eigenstate $m$ of an ideal lossless isolated stationary optical microresonator varies in time as $\exp(-i\omega_m t)$ with eigenfrequency $\omega_m$. The eigenstates of such resonator cannot attenuate or grow and, thus, its eigenfrequencies $\omega_m$ cannot be complex valued. Situation is different for the lossless *periodically modulated* systems. The behaviour of such systems is described by differential equations (e.g., Maxwell and wave equations) whose parameters periodically depend on time. It is well known that solutions of such differential equations can be stable as well as unstable [1, 2]. In the case of stability, the discrete set of quasi-states and corresponding quasi-eigenfrequencies can be introduced. In the case of instability, the eigenfrequencies of quasi-states of an ideal lossless optical resonator belong to the continuous spectrum [1-5].

In classical mechanics, resonant oscillations of a periodically modulated resonator, which infinitely grow with time in the absence of losses, become finite in their presence [6]. Similarly, light in a modulated resonator, being unstable in the absence of losses, qualitatively changes its behaviour in their presence. For relatively small losses, the eigenstate of an unmodulated microresonator attenuates as $\exp(-i\omega_m t - \alpha_m t)$ where the imaginary part $\alpha_m$ of eigenfrequency is introduced [7]. The eigenfrequency $\omega_m$ of a lossless optical microresonator, which parameters are *periodically modulated* with frequency $\omega_p$, shifts to $\omega'_m$ and splits into a comb with frequencies $\omega'_m + n\omega_p$ numerated by integer $n$ (see, e.g., [8]). A practically important and still not fully understood problem is concerned with the behaviour of spectrum of *parametrically modulated resonators with losses* which are introduced internally and due to the leakage to the input-output waveguides. Of special interest is the case of resonant excitation, when the modulation frequency $\omega_p$ coincides with the free spectral range (FSR) of the microresonator. The interest to this problem has been boosted by recent experimental demonstrations of optical frequency combs generated by parametrically excited microresonators [9, 10] following significant previous work in the field (see [9-16] and references therein). In particular, it has been discovered, both theoretically and experimentally, that, in addition to splitting into $\omega_p$-periodic combs, the presence of losses leads to additional fine structure of microcomb resonances as a function of the input light frequency [9, 10].

Here we develop the semiclassical theory of parametrically excited microcombs which assumptions fit the common experimental conditions. We show that, in the semiclassical approximation, calculation of the transmission amplitude exhibiting optical frequency comb (OFC) is reduced to linear functional equations which can be solved analytically. The developed theory accounts for multi-quantum transitions between the microresonator eigenstates as well as for the spatial distribution of modulation (SDM). We determine the complex-shaped profile of individual comb transmission resonances formed by the interplay between the fine splitting of eigenfrequencies proportional to the modulation amplitude and their broadening determined by the attenuation of light. We express this profile through the spatial and temporal dependencies of modulation parameters and optimize the SDPM to minimize the consumption of power required for the OFC generation.

As an application of the developed theory, we consider racetrack microresonators (RTMs) and SNAP bottle microresonators (SBMs) illustrated in Fig. 1. Two critical benefits of the SBMs compared to RTMs are revealed. We show that the effect of parametric excitation is significantly enhanced in SBMs compared to RTMs with similar dimensions due to the slow propagation speed of whispering gallery modes (WGMs) along the SBM axis compared to the speed of light along the RTM waveguide. Consequently, we show that, to generate an OFC with the same repetition rate, the dimensions of an SBM can be much smaller than the dimensions of an RTM.

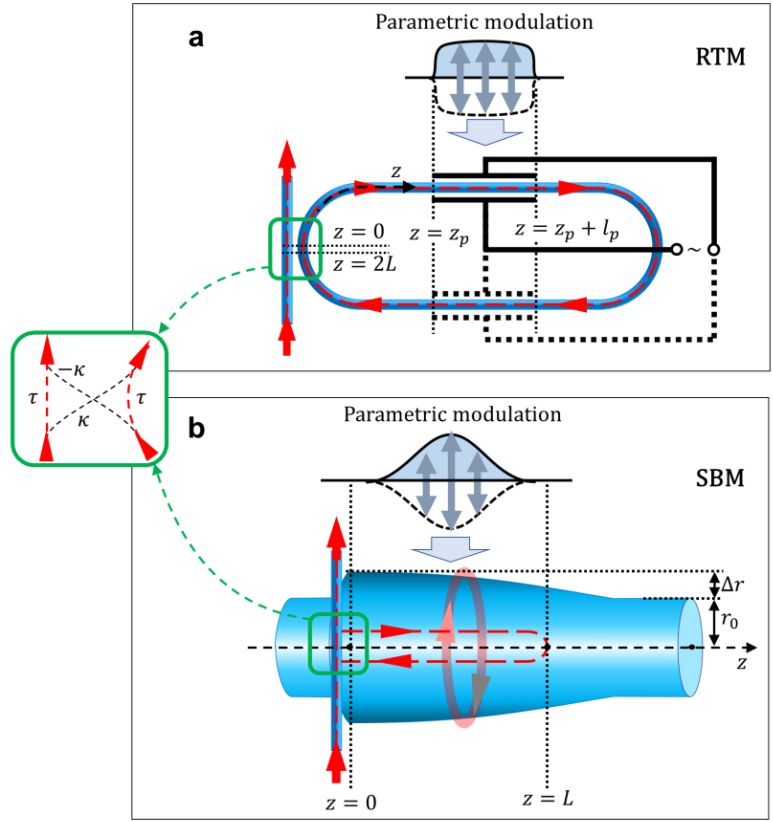

**Fig. 1.** (a) Parametrically modulated RTM. (b) Parametrically modulated SBM. Inset: Magnified coupling region between the input-output waveguide and microresonators.

## 2. Racetrack microresonator (RTM)

We consider now an RTM coupled to an input-output waveguide illustrated in Fig. 1(a). We start with the consideration of the one-dimensional propagation of linearly polarized light localized along the RTM waveguide away from the coupling region. This propagation is described by the Maxwell equations $\frac{\partial E_y}{\partial z} = -\frac{c^{-1}\partial H_x}{\partial t}$, $\frac{\partial H_x}{\partial z} = -\frac{c^{-1}\partial D_y}{\partial t}$. Here $c$ is the speed of light and the dielectric displacement $D_y(z,t)$ is related to the electric field $E_y(z,t)$ as $D_y(z,t) = n^2(z,t)E_y(z,t)$ where $n(z,t)$ is the effective refractive index. Combining these equations, we arrive at the wave equation:

$$\frac{\partial^2 E}{\partial z^2} - \frac{1}{c^2}\frac{\partial^2 D}{\partial t^2} = 0, \quad E(z,t) = D(z,t)/n^2(z,t). \tag{1}$$

For brevity, in Eq. (1) and below we omit the coordinate indices of $E$ and $D$. We solve Eq. (1) for the refractive index $n(z,t)$ periodically modulated in time with frequency $\omega_p$:

$$n(z,t) = n_0 + i\eta + \Delta n_{p0}(z) + \Delta n_{p1}(z)\cos(\omega_p t) \tag{2}$$

Here $n_0$ is the original constant refractive index of the RTM waveguide and $\eta \ll n_0$ determines small material and scattering losses. In agreement with common experimental conditions, we assume that the modulation frequency $\omega_p$ is much smaller than the frequency of input light $\omega_{in}$, i.e., $\omega_p \ll \omega_{in}$, and the amplitudes of the refractive index perturbations $\Delta n_{pj}(z), j = 1,2$, are small compared to the original waveguide refractive index, $|\Delta n_{pj}(z)| \ll n_0$. In addition, these amplitudes are assumed to vary in space slow compared to the oscillations of light determined by its characteristic propagation constant $\beta = \frac{\omega_{in} n_0}{c}$, i.e., $\left|\frac{\partial \Delta n_{pj}}{\partial z}\right| \ll \beta|\Delta n_p|$ [17]. In the approximation considered, the spectral bandwidth of solution of Eq. (1) (and, therefore, the OFC bandwidth) should be much smaller than the frequency of the input light $\omega$. For the modulation induced by linear electro-optics Pockels effect (like, e.g., in lithium niobate [9, 10, 14]), $\Delta n_{p1}(z)\cos(\omega_p t)$ in Eq. (2) is proportional to the applied voltage, $U\cos(\omega_p t)$, and $\Delta n_{p0}(z)$ can be set to zero. For the modulation induced by electrostriction and Kerr effect, $\Delta n_{p0}(z) + \Delta n_{p1}\cos(\omega_p t)$ is proportional to the applied voltage squared, $(U\cos(\frac{\omega_p t}{2}))^2$, which contributes both to $\Delta n_{p1}(x)$ and $\Delta n_{p0}(z)$.

Outside the coupling region indicated by the rectangle in Fig. 1(a), the field $E(x,t)$ propagating along the microresonator waveguide is determined by Eqs. (A3) and (A6) of Appendix A. Here, we describe the waveguide coupling the commonly used transfer matrix approach (see e.g., [18, 19]). For determinacy, we assume that light propagates into the positive direction along the RTM waveguide axis $z$. As illustrated in Fig. 1(a), the coordinate $z = 2L$ corresponds to the beginning of the coupling region and coordinate $z = 0$ corresponds to its end. The monochromatic light launched into the input-output waveguide is set to $E_{in}(t) = \exp(-i\omega_{in}t)$. Then, the output light $E_{out}(t)$ is determined by the equation:

$$\begin{pmatrix} E_{out}(t) \\ E(0,t) \end{pmatrix} = S \begin{pmatrix} E_{in}(t) \\ E(2L,t) \end{pmatrix}, \quad S = \begin{pmatrix} \tau & \kappa \\ -\kappa & \tau \end{pmatrix}. \tag{3}$$

For the lossless coupling assumed below, matrix $S$ introduced in this equation is the unitary S-matrix. Without loss of generality, we assume that the elements of this matrix, $\tau$ and $\kappa$, are real [20]. Then, the condition of unitarity yields $\tau^2 + \kappa^2 = 1$. As shown in Appendix C, substitution of the solution of Eq. (1) at $z = 0$ and $z = L$ into Eq. (3) leads to the functional equation whose solution is found analytically in Appendix B. As the result, calculations detailed in Appendix C yield the comb spectral components $E_m$ of $E_{out}(t)$:

$$E_{out}(t) = \sum_{m=-\infty}^{\infty} E_m \exp\left[i\left(m\omega_p - \omega_{in}\right)t\right],$$
$$E_m = \tau\delta_{0m} - \kappa^2 \exp\left[im\left(\frac{\pi}{2} - \frac{\omega_p T}{2} + \arg(\Omega_{p1})\right)\right] \tag{4}$$
$$\times \sum_{n=0}^{\infty} \tau^n J_m\left(\sigma_{n+1}|\Omega_{p1}|\right)\exp\left[(n+1)\left(-\frac{im}{2}\omega_p T + i\omega_{in}T - \frac{\eta}{n_0}\omega_{in}T + i\Omega_{p0}\right)\right].$$

Here $\delta_{0m}$ is the Kronecker delta, $T$ is the time of circulation of light along the microresonator circumference $2L$ [21],

$$T = \frac{2n_0 L}{c} = \frac{2\beta_0 L}{\omega_0}, \tag{5}$$

$\Omega_{p0}$ and $\Omega_{p1}$ are the modulation parameters,

$$\Omega_{p0} = \frac{\omega_{in}}{c}\int_0^{2L} dz \Delta n_{p0}(z), \quad \Omega_{p1} = \frac{\omega_{in}}{c}\int_0^{2L} dz \Delta n_{p1}(z)\exp\left(\frac{i\omega_p n_0 z}{c}\right), \tag{6}$$

and

$$\sigma_n = \frac{\sin\left(\frac{n}{2}\omega_p T\right)}{\sin\left(\frac{1}{2}\omega_p T\right)} \tag{7}$$

is the resonant index. Results similar to Eqs. (4) were previously obtained based on the lumped modulation parameter model [9, 12, 22]. Here we elucidate and clarify those results by presenting the expressions for the lumped parameters $\Omega_{p0}$ and $\Omega_{p1}$ (Eq. (6)) through the spatial variation of modulated refractive index. Eq. (4) is more compact compared to those obtained previously since it contains only a single sum over the turns of light in the microresonator rather than a double sum in Refs. [12, 22]. In previous considerations, parameter $\Omega_{p0}$ proportional to the time-independent component of the external perturbation was usually set to zero, while the value of modulation index $\Omega_{p1}$ was often set equal to a real empirical constant [9, 10, 12, 14, 22]. Eq. (4) shows that the magnitude of combs $|E_m|$, except for the central resonance with $m = 0$, does not depend on the phase of $\Omega_{p1}$ and, thus, the approximation of real $\Omega_{p1}$ can be justified for $m \neq 0$. Eqs. (4)-(6) allow to investigate and maximize the comb amplitudes $E_m$ by optimization of the modulation index $\Omega_{p1}$, which is a functional of the refractive index SDM $\Delta n_{p1}(x)$. Remarkably, the introduction of transmission effective modulation index $\sigma_n|\Omega_{p1}|$ (the argument of the Bessel's functions in Eq. (4)) allows us to simplify the expressions for the comb amplitudes as compared to those derived in Ref. [9, 22].

The cases of major interest described by Eqs. (4)-(7) correspond to the *resonant or adiabatic modulation* when many terms contribute into the sum over $n$. To this end, the phases of these terms must be close to each other, i.e., satisfy the condition of constructive interference. In particular, the attenuation due the losses inside the microresonator and coupling should be small, i.e., $\kappa = \sqrt{1-\tau^2} \ll 1$ and $\frac{2\eta\omega L}{c} \ll 1$, as assumed hereafter. The resonant or adiabatic excitation can take place only at or close to the conditions $\omega_p T \to 2\pi N, N = 0, 1, 2, ...,$ which includes the adiabatic case $\omega_p T \to 0$. At these conditions, we have

$$\sigma_{n+1} = (-1)^{Nn}(n+1). \tag{8}$$

Then, as shown in Appendix C, we find

$$E_{out}(t) = E_{in}(t)\frac{\tau - U(t)}{1-\tau U(t)},$$
$$U(t) = \exp\left[i(\omega_{in}-\omega_q)T - \frac{2\eta\omega_{in}L}{c} + i|\Omega_{p1}|\cos\left(\omega_p t + \arg(\Omega_{p1})\right)\right] \tag{9}$$

Here, we introduced the RTM eigenfrequencies $\omega_q$ corresponding to vanishing modulation, $\Omega_{p1} = 0$, which are defined by the quantization rule

$$\omega_q T + \Omega_{p0} = 2\pi(s_0 + q), \tag{10}$$

where integers $s_0 \gg 1$ and $q \ll s_0$. The eigenfrequencies defined by this equation coincide which those found in Appendix D for a standing along RTM. For the resonant transmission considered below, we chose the eigenfrequency $\omega_0$ close to the input frequency $\omega_{in}$ so that $|\omega_{in} - \omega_0| \ll |\omega_1 - \omega_0|$ and assume that $q \ll s_0$ to satisfy the conditions of our approximation $|\omega_q - \omega_0| \ll \omega_0$. The result similar to Eq. (9) was previously obtained in Refs. [9, 22] based on the lumped modulation parameter model with a real modulation index $\Omega_{p1}$. For the weak modulation $|\Omega_{p1}| \ll 1$, Eq. (9) coincides with that obtained in Ref. [10]. In the latter case, we expand the exponent in the expression for $U(t)$ in Eq. (9) keeping only the first-order term in $|\Omega_{p1}|$ and, thus, take into account only the single quantum transitions with acquisition or loss of frequency $\omega_p$ during a single roundtrip of light along the RTM waveguide.

Conditions when the amplitudes $|E_m|$ of comb lines with large numbers $m \gg 1$ are resonantly enhanced are of the major practical interest. For very small deviations from the parametric resonance,

$$\omega_p T = 2\pi N + \Delta\omega_p T, \quad |\Delta\omega_p T| \ll \alpha, \quad \alpha = \frac{\kappa^2}{2} + \frac{2\eta\omega_{in}L}{c}, \tag{11}$$

asymptotic calculations detailed in Appendix E yield the OFC amplitude variation near the resonant frequency $\omega_q = \omega_0 + \frac{2\pi q}{T}$:

$$|E_m| \cong \kappa^2 |\Omega_{p1}|^{-1/2} \mathcal{E}_m, \quad \mathcal{E}_m = |1 - \Lambda_m^2|^{-1/2} \left|\Lambda_m - i\sqrt{1 - \Lambda_m^2}\right|^{|m|},$$

$$\Lambda_m = \Delta\omega_{T\Omega} + i\alpha_\Omega, \quad \Delta\omega_{T\Omega} = \frac{(\omega_{in} - \omega_q)T - \frac{1}{2}m\Delta\omega_p T}{|\Omega_{p1}|}, \quad \alpha_\Omega = \frac{\alpha}{|\Omega_{p1}|}. \tag{12}$$

Here we introduce the normalized comb line amplitude $\mathcal{E}_m = |E_m| \kappa^{-2} |\Omega_{p1}|^{1/2}$, and also dimensionless frequency deviation $\Delta\omega_{T\Omega}$ and full roundtrip attenuation $\alpha_\Omega$ measured in units of modulation index $|\Omega_{p1}|$. At the exact parametric resonance, when $\Delta\omega_p T = 0$, and $\omega_p T = 2\pi N, N > 1$, parameter $\Lambda_m$ does not depend on OFC line number $m$. In this case, Eq. (12) coincides with that obtained in Ref. [10] under the condition of weak modulation index $|\Omega_{p1}| \ll 1$. From Eq. (12), the slowest variation of the OFC power with the line number $m$ is achieved either at the smallest possible $\Lambda$, i.e., for $\omega_{in} = \omega_s$, or for $\Lambda \cong 1$. Alternatively, in the adiabatic case, $N = 0$ and $\omega_p T = \Delta\omega_p T$, and for a small deviation of the modulation frequency $\omega_p$ from the parametric resonance condition, parameter $\Lambda_m$, even being fixed at its smallest value at $m = 0$, will slowly move out of resonance for larger $m$.

Numerical comparison of the comb amplitudes found from Eq. (12) and Eq. (4) demonstrates the remarkable accuracy of Eq. (12) both for weak modulation, $|\Omega_{p1}| \ll 1$ [10, 14], and strong modulation, $|\Omega_{p1}| \sim 1$ [9, 22]. As an example, Figs. 2(a)-(c) compare the behaviour of amplitudes $|E_m|$ as a function of dimensionless frequency variation $\Delta\omega_{T\Omega}$ of the input light for the case of exact resonance, $\omega_p T = 2\pi$, modulation index $|\Omega_{p1}| = 1$, coupling parameter $\kappa = 0.04$, and different attenuation parameters $\alpha = 0.05, 0.01$, and $0.002$. In agreement with the condition of validity of Eq. (12), much better agreement, which also holds for $|\Omega_{p1}| \gg 1$, takes place in the vicinity of the maxima of the comb amplitudes, where $||\Lambda| - 1| \ll 1$. For the combs with numbers $m$ smaller than $m_{inf}$ corresponding to the inflection condition, $m < m_{inf}$, the plots in Figs. 2(a)-(c) exhibit two peaks with maxima indicated by circles and one minimum. For greater numbers $m > m_{inf}$, these plots have a single maximum. From Eq. (12), two conditions, $\Lambda \to 0$ and $\Lambda \to \pm 1$, correspond to the flatten frequency combs since then $|\Lambda - i\sqrt{1 - \Lambda^2}| \to 1$ and the comb peaks amplitude becomes slow function of $m$. The first condition corresponds to two maxima peak values of the plots in Figs. 2(a)-(c) at $m < m_{inf}$, while the second condition corresponds to the single minimum of these plots at $m > m_{inf}$. For small deviation from the parametric resonance satisfying Eq. (11), Eq. (12) remains sufficiently accurate as well, as illustrated in Fig. 3(d) for $\Delta\omega_p T = 0.001$. It is seen from this figure that the non-zero $\Delta\omega_p T$ leads to the small asymmetry of the comb line power plot with respect to $\Delta\omega_{T\Omega} = 0$.

For applications, it is usually important to evaluate the comb line amplitude as a function of $m$ for the fixed frequency. However, it is also interesting to look at the dependence of the comb line power maxima as a function of their number shown in Fig. 2(e). It is seen that, in the dB scale, this dependence on $m$ is nonlinear at $m < m_{inf}$ and linear at $m > m_{inf}$ following Eq. (12).

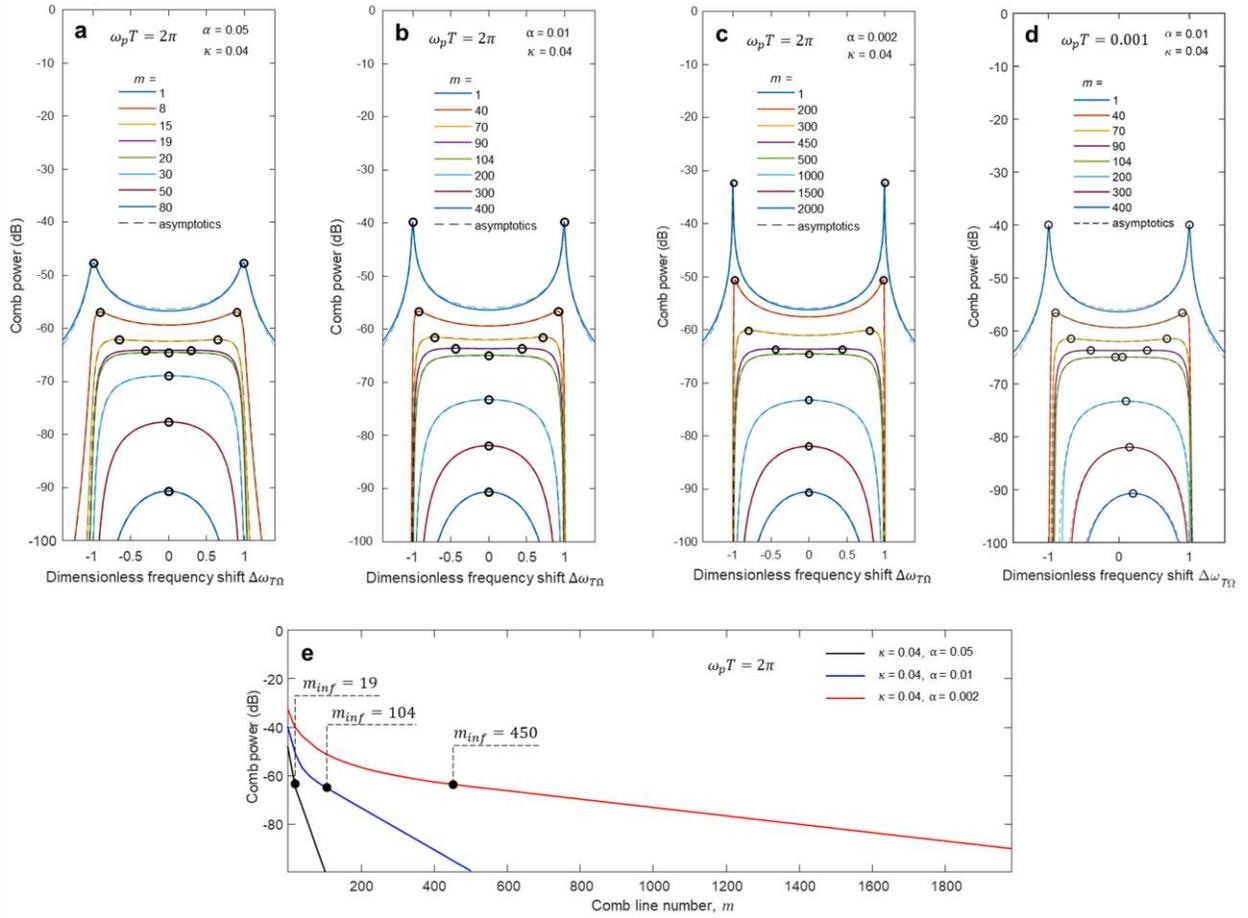

**Fig 2.** (a)-(c) The behaviour of comb line power in dB, $20\log(|E_m|)$, as a function of dimensionless frequency variation $\Delta\omega_{T\Omega}$ of the input light for the case of parametric resonance, $\omega_p T = 2\pi$, modulation index $|\Omega_{p1}| = 1$, coupling parameter $\kappa = 0.04$, and full round trip attenuation (a) $\alpha = 0.05$, (b) $\alpha = 0.01$ and (c) $\alpha = 0.002$. (d) The comb line power for the adiabatic modulation $\omega_p T = 0.001$; here other parameters are the same as in (b). In (a)-(d), circles indicate the maxima of plots. (e) The behaviour of the maximum comb line power as a function of $m$ for $\kappa = 0.04$, and $\alpha = 0.05$, $0.01$, and $0.002$.

## 3. SNAP bottle microresonator (SBM)

SNAP microresonator structures are fabricated at the surface of an optical fibre by nanoscale deformation [23]. Several approaches for fabrication of these resonators developed to date achieve the unprecedented subangstrom precision [23-30]. Commonly, SNAP microresonators are fabricated from the silica fibres, which parameters (refractive index distribution and dimensions) can be varied in time by applying strong modulated laser field. The frequency of this field is supposed to be separated from a weaker field generating OFC due to the induced modulation of the SBM parameters and the OFC spectrum itself. For pure silica fibres, modulation can be induced by the combined nonlinear Kerr and radiation pressure effects [31]. Generally, SBMs can be introduced at the surfaces of fibres fabricated of materials other than pure silica. These fibres include fibres fabricated of dopped silica and silica fibres integrated with other materials [32, 33, 34], as well as accurately polished [10] fibres fabricated of highly nonlinear materials such as lithium niobate [35, 36]. Modulation of the SBMs fabricated at the surfaces of these fibres can be induced by the applied electromagnetic field through Pockels, Kerr and radiation pressure effects, as well as their combinations, and PZT-induced mechanical vibrations [37, 38].

The SNAP bottle microresonator (SBM) considered in this section is illustrated in Fig. 1(b). Light is localized in an SBM in the form of whispering gallery modes (WGMs) which circulate near the fibre surface and slowly propagate along its length $z$. Thanks to dramatically small variation of the SBM parameters along the axial coordinate $z$, the expression for a WGM propagating in an SBM is factorized as $\mathcal{E}_{mp}(\varphi, r, z, t) = \exp(im\varphi) Q_{mp}(\varrho) E_{mp}(z)$ where $(z, \varrho, \varphi)$ are the cylindrical coordinates. The WGM slowness is ensured by the condition that its frequency $\omega$ is close to a fibre cutoff frequency (CF) $\omega_{cut}(z,t)$, i.e., $|\omega - \omega_{cut}(z,t)| \ll \omega$. The CF varies along the fibre length and changes in time due to the modulation of SBM parameters. First, we introduce the original CF of the unperturbed SBM, $\omega_{cut}^0(z)$ and its relatively small variation $\Delta\omega_{cut}^0(z)$ so that

$$\Delta\omega_{cut}^0(z) = \omega_{cut}^0(z) - \omega_0. \tag{13}$$

Here frequency $\omega_0$ is assumed to be adjacent to the CF, and the CF variation $\Delta\omega_{cut}^0(z)$ is assumed to be small enough, $|\Delta\omega_{cut}^0(z)| \ll \omega_0$. The full time-dependent CF variation of the SBM is the sum of original stationary CF variation $\Delta\omega_{cut}^0(z)$, which determines the SBM shape, much smaller variation $\Delta\omega_{p0}(z) + \Delta\omega_{p1}(z)\cos(\omega_p t)$ introduced by parametric modulation, and complex parameter $-i\gamma$ determined by the material losses:

$$\Delta\omega_{cut}(z, t) = \Delta\omega_{cut}^0(z) + \Delta\omega_{p0}(z) + \Delta\omega_{p1}(z)\cos(\omega_p t) - i\gamma, \tag{14}$$

where $|\Delta\omega_{p0}(z,t)|, |\Delta\omega_{p1}(z,t)|, \gamma \ll |\Delta\omega_{cut}^0(z)|$. Under these conditions, the $z$-dependence of the WGM field, $E_{mp}(z)$, is described by the Schrödinger equation [39]:

$$i\frac{\partial E}{\partial t} + \frac{\chi}{2}\frac{\partial^2 E}{\partial z^2} - \omega_{cut}(z,t)E = 0, \quad \omega_{cut}(z,t) = \omega_0 + \Delta\omega_{cut}(z,t), \quad \chi = \frac{c^2}{n_0^2 \omega}. \tag{15}$$

Here the CF $\omega_{cut}(z,t)$ is a sufficiently smooth function of $z$ everywhere except, possibly, at the microresonator edges. Equations (14) and (15) were previously used in [40] to determine the OFC spectrum for the model of time-dependent harmonic oscillator. A more general numerical analysis of the OFC generation based on these equations was performed in [41]. Here we assume that the CF variation $\Delta\omega_{cut}^0(z)$ is sufficiently large and/or expanded along the axial direction to support multiple eigenmodes localized in the SBM as required for the generation of multiple comb lines. We introduce the propagation constant $\beta(z, \omega)$ and propagation time $\tau(z, \omega)$ in the unmodulated SBM:

$$\beta(z, \omega) = \sqrt{\frac{2}{\chi}(\Delta\omega - \Delta\omega_{cut}^0(z))}, \quad \tau(z, \omega) = \frac{1}{\chi}\int_0^z \frac{dz}{\beta(z, \omega)}, \quad \Delta\omega = \omega - \omega_0. \tag{16}$$

Under the assumptions made, solution of Eq. (15) away from the turning points can be found in the semiclassical approximation [42] (see Appendix F). This solution allows us to determine the optical frequency comb spectrum of the modulated SBM using the approach similar to that developed in the previous section for the RTM. In particular, the eigenmodes and eigenfrequencies of a standing along SBM are presented in Appendix G. For the stationary CF, $\Delta\omega_{p1}(z) = 0$, the frequency eigenvalues of the SBM are determined by the semiclassical quantization rule similar to Eq. (G5) of Appendix G:

$$2\int_0^L \beta(z, \omega_q)dz + \tilde{\Omega}_{p0} = 2\pi(s_0 + q + \varsigma),$$
$$\tilde{\Omega}_{p0} = \frac{2}{\chi}\int_0^L \frac{\Delta\omega_{p0}(z)}{\beta(z, \omega_0)}dz. \tag{17}$$

Here $s_0 \gg 1$ and $q$ are integer quantum numbers and $\varsigma \sim 1$ is the phase shift which appears due to the reflection of semiclassical solutions of Eq. (15) from the turning points. Similar to Eq. (10), we set $\omega_0$ equal to an eigenfrequency corresponding to $s = s_0$ and consider OFCs localized near $\omega_0$. From Eq. (16), the full roundtrip time of WGM circulation between the SBM axial turning points $z = 0$ and $z = L$ at $\omega = \omega_0$ is

$$T = 2\tau(L, \omega_0). \tag{18}$$

We assume now that an SBM is coupled to a transversely oriented input-output waveguide positioned close to the left SBM edge as illustrated in Fig. 1(b). The configuration shown in Fig. 1(b) is similar to that experimentally investigated in Ref. [24]. In the right-hand side of the coupling region, the solution of Eq. (15) can be presented as a sum of solutions propagating into the positive and negative direction along the RTM, $E^{\rightarrow}(z,t) + E^{\leftarrow}(z,t)$. As in the previous section, we assume that the modulation is absent in a certain vicinity to the right of the coupling region. In this vicinity, the boundary condition is similar to that of the coupled RTM considered in Section 2:

$$\begin{pmatrix} E_{out}(t) \\ E^{\rightarrow}(0,t) \end{pmatrix} = S \begin{pmatrix} E_{in}(t) \\ E^{\leftarrow}(0,t) \end{pmatrix}, \quad S = \begin{pmatrix} \tau & \kappa \\ -\kappa & \tau \end{pmatrix}. \tag{19}$$

Here $E_{in}(t) = \exp(-i\omega_{in}t)$ and $E_{out}(t) = \exp(-i\omega_{in}t) E_{out}^0(t)$ are the input and output fields in the waveguide. At the right edge of the SBM, the boundary condition for the solution $E^{\rightarrow}(z,t) + E^{\leftarrow}(z,t)$ corresponds to the semiclassical condition of its evanescent decay for $z > L$ [7]:

$$E^{\rightarrow}(z,t)\big|_{z \nearrow L} = \exp\left(-\frac{i\pi}{2}\right) E^{\leftarrow}(z,t)\big|_{z \nearrow L} \tag{20}$$

Solution of Eq. (15), $E^{\rightarrow}(z,t) + E^{\leftarrow}(z,t)$, with boundary conditions determined by Eqs. (19) and (20) detailed in Appendix I yields the expression for the output field:

$$\begin{aligned}
E_{out}(t) &= \sum_{m=-\infty}^{\infty} E_m \exp\left[i(m\omega_p - \omega_{in})t\right], \\
E_m &= \tau\delta_{0m} - \kappa^2 \exp\left(\frac{i\pi m}{2}\right) \times \\
&\sum_{n=0}^{\infty} \tau^n J_m(\tilde{\sigma}_{n+1}) \exp\left[(n+1)\left(-\frac{im}{2}\omega_p T + 2i\int_0^L \beta(z,\omega_0)dz + i\tilde{\Omega}_{p0} - \gamma T - \frac{i\pi}{2}\right)\right],
\end{aligned} \tag{21}$$

where the modulation induced phase shift $\tilde{\Omega}_{p0}$ and modulation parameter $\tilde{\Omega}_{p1}$ are determined by equations

$$\tilde{\Omega}_{p0} = \frac{2}{\chi}\int_0^L \frac{\Delta\omega_{p0}(z)}{\beta(z,\omega_0)}dz, \quad \tilde{\Omega}_{p1} = \frac{2}{\chi}\int_0^L \Delta\omega_{p1}(z)\cos\left(\omega_p\left(\frac{T}{2}-\tau(z,\omega_0)\right)\right)\frac{dz}{\beta(z,\omega_0)}, \tag{22}$$

and

$$\tilde{\sigma}_n = \frac{\sin\left(\frac{n}{2}\omega_p T\right)}{\sin\left(\frac{1}{2}\omega_p T\right)}\left|\tilde{\Omega}_{p1}\right|. \tag{23}$$

In full analogy to the derivation of Eq. (12) for the RTM, the asymptotic of amplitudes $|E_m|$ at exact parametric resonance $\omega_p T = 2\pi N, N = 0, 1, 2, \ldots$, is

$$\begin{aligned}
|E_m| &\cong \kappa^2 \left|\tilde{\Omega}_{p1}\right|^{-1/2} \tilde{\mathcal{E}}_m, \quad \mathcal{E}_m = \left|1-\tilde{\Lambda}^2\right|^{-1/2}\left|\tilde{\Lambda} - i\sqrt{1-\tilde{\Lambda}^2}\right|^{|m|}, \\
\tilde{\Lambda} &= \Delta\omega_{T\Omega} + i\alpha_\Omega^+, \quad \Delta\omega_{T\Omega} = \frac{(\omega_{in}-\omega_q)T}{\left|\tilde{\Omega}_{p1}\right|}, \quad \tilde{\alpha}_\Omega^+ = \frac{\alpha^+}{\left|\tilde{\Omega}_{p1}\right|}, \quad \alpha^+ = \frac{\kappa^2}{2} + \frac{\eta\omega_{in}T}{n_0}.
\end{aligned} \tag{24}$$

Here $\tilde{\mathcal{E}}_m = |E_m| \kappa^{-2} |\tilde{\Omega}_{p1}|^{1/2}$ is the normalized comb line amplitude, and $\Delta\omega_{T\Omega}$ and $\tilde{\alpha}_\Omega^+$ are dimensionless frequency deviation and full roundtrip attenuation measured in units of modulation index $|\tilde{\Omega}_{p1}|$. The validity of Eq. (24) is limited by the validity of the semiclassical solution of Eq. (15). In particular, it assumes that the OFC bandwidth is small compared to the bandwidth of the SBM discrete spectrum.

### 4. SBM vs. RTM

We are now ready to compare the OFC generation characteristics of the RTM and SBM. Due to the similarity of expressions for the OFC spectra of the RTM and SBM (compare Eqs. (4), (6), (7), (10), (12) and Eqs. (17), (21)-(24)), the major difference between these characteristics is included in the expressions for modulation parameters $\Omega_{p1}$ and $\tilde{\Omega}_{p1}$ defined by Eqs. (6) and (22). The most interesting for applications are the cases of resonant and adiabatic parametric modulation when

$$\omega_p T = 2\pi N, \quad N = 0, 1, 2, \ldots \tag{25}$$

where $N = 0$ corresponds to the adiabatic modulation, $\omega_p T \ll 2\pi$. For $N \geq 1$, the microresonator length enabling the resonant modulation is proportional to $N$, e.g., for a RTM $L = \pi N c / n_0 \omega_p$. In the adiabatic case, the OFC repetition rate is much smaller than the microresonator FSR, $\omega_p \ll \frac{2\pi}{T}$, and the microresonator eigenfrequency dispersion is irrelevant. In contrast, in the resonant case, $N \geq 1$, the equispacing of the eigenfrequencies, which correspond to the frequencies of the generated comb lines, is critical. The semiclassical perturbation theory used above assumes that the FSR $2\pi/T$ is constant (i.e., the eigenfrequency dispersion is small) with sufficient accuracy. In the case of RTM, the latter assumption suggests that the value of modulation indices, $\Omega_{p1}$ and $\tilde{\Omega}_{p1}$, rather than the dispersion determines the OFC bandwidth. In the case of SBM, the dispersion can be minimized for microresonators with parabolic, semi-parabolic [24] as well as completely different from parabolic [39] CF variation. Increasing the SBM dimensions allow to reduce its axial FSR and dispersion, as e.g., in long rectangular SBM [29]. However, the OFC bandwidth may be limited by the relatively narrow bandwidth of the SBM discrete spectrum (see the discussion below).

Hereafter, we distinguish the parameters of an RTM and an SBM (where necessary to avoid confusion) by adding "~" at the top of the SBM parameters. For example, the roundtrip time and refractive index are denoted as $T$ and $n_0$ for an RTM and as $\tilde{T}$ and $\tilde{n}_0$ for an SBM. Our analysis in the previous sections considered the refractive index SDM $\Delta n_{p1}(z)$ for an RTM and the CF modulation $\widetilde{\Delta\omega}_{p1}(z)$ for an SBM. To compare the effects of these modulations, we express the amplitude $\widetilde{\Delta\omega}_{p1}(z)$ of an SBM through the amplitude of its effective refractive index modulation $\widetilde{\Delta n}_{p1}(z)$ and effective radius variation $\widetilde{\Delta r}_0(z)$ with the rescaling relation:

$$\frac{\Delta\omega_{p1}(z)}{\tilde{\omega}_0} = -\frac{\Delta n_{p1}(z)}{\tilde{n}_0} = -\frac{\Delta r_0(z)}{r_0}. \tag{26}$$

We define the *OFC-equivalent* microresonators as microresonators generating similar OFCs. Provided that the generated OFC bandwidth is not limited by the microresonators' dispersion and by the SBM spectral bandwidth, it follows from the developed theory that the *resonantly modulated RTM and SBM are OFC-equivalent if they generate OFCs with the same modulation indices, $\Omega_{p1} = \tilde{\Omega}_{p1}$, central resonance frequency $\omega_0$, and repetition rate $\omega_p$.* Excluding the adiabatic modulation, the latter condition, together with the resonant condition of Eq. (25), determines the roundtrip time $T = \frac{2\pi N}{\omega_p}$, which should be the same for the OFC-equivalent RTM and SBM as well.

It is instructive, first, to consider the case of the uniform SDMs, when $\Delta n_{p1}(z) = \Delta n_{p1}^0$ for the RTM and $\widetilde{\Delta\omega}_{p1}(z) = \widetilde{\Delta\omega}_{p1}^0$ for the SBM. Then, at the exact modulation resonance, $\omega_p T = 2\pi N$, $N \geq 1$, calculations yield $\Omega_{p1} = \tilde{\Omega}_{p1} = 0$. Thus, the uniform SDM does not generate OFCs (for comparison, see Appendix D describing a standing along RTM). This is simply explained by the fact that, due to the orthogonality of the microresonator eigenstates, the matrix elements of a spatially uniform perturbation at the different eigenstates are zero. In contrast, for the adiabatic modulation, $\omega_p \ll \frac{2\pi}{T}$ and $\tilde{\omega}_p \ll \frac{2\pi}{\tilde{T}}$, we have from Eqs. (6), (22), and (26):

$$\Omega_{p1} = \frac{\Delta n_{p1}^0}{n_0} T \omega_0, \quad \tilde{\Omega}_{p1} = \frac{\widetilde{\Delta n}_{p1}^0}{\tilde{n}_0} \tilde{T} \tilde{\omega}_0. \tag{27}$$

From this equation, for the same refractive indices, $n_0 = \tilde{n}_0$, modulation amplitudes, $\Delta n_{p1}^0 = \widetilde{\Delta n}_{p1}^0$, and resonant frequencies, $\omega_0 = \tilde{\omega}_0$, the SBM modulation index $\tilde{\Omega}_{p1}$ can be much greater than the RTM modulation index $\Omega_{p1}$ due to the slow WGM propagation in an SBM leading to a much greater roundtrip time $\tilde{T} \gg T$.

Generally, in contrast to RTMs, light propagating in SBMs experiences multiple transverse circulations around the SBM surface before completing the roundtrip along its axis $z$. Consequently, the axial propagation speed of WGMs along the SBM axis is much slower than the speed of light along the RTM waveguide. Another important difference is that WGMs propagating into the positive and negative directions, $E^{\rightarrow}(z,t)$ and $E^{\leftarrow}(z,t)$, along the SBM axis spatially overlap and experience the same local effect of the parametric modulation. For this reason, here we will compare the values of $\Omega_{p1}$ and $\tilde{\Omega}_{p1}$ assuming that the SDM of refractive index in the RTM is $L/N$-translationally antisymmetric, i.e., $\Delta n_{p1}(z + (L/N)) = -\Delta n_{p1}(z)$. This condition also corresponds to the minimum consumption power (see the discussion below). Then, Eq. (6) for $\Omega_{p1}$ at the resonance condition of Eq. (25) is transformed into

$$\Omega_{p1} = 2N \frac{\omega_0}{c} \int_0^{L/N} dz \Delta n_{p1}(z) \exp\left(\frac{i\pi N}{L} z\right). \tag{28}$$

To understand the major difference between OFC generated by the RTMs and SBMs, we assume that the SBM is rectangular (see Appendix H), i.e., its CF is constant within the microresonator length, $\omega_{cut}^0(z) = \tilde{\omega}_c^0$ [43]:

$$\tilde{\beta}(z, \tilde{\omega}_0) = \tilde{\beta}_0 = \frac{\tilde{n}_0}{c}\sqrt{2\tilde{\omega}_0 \Delta \tilde{\omega}_0}, \quad \Delta \tilde{\omega}_0 = \tilde{\omega}_0 - \tilde{\omega}_c^0. \tag{29}$$

Here $\widetilde{\Delta\omega}_0$ is the difference between the unperturbed CF of the SBM and its quasi-eigenfrequency $\tilde{\omega}_0$ defined by Eq. (17). Then, we find for the roundtrip time for the SBM of length $\tilde{L}$:

$$T = \frac{\tilde{n}_0}{c}\sqrt{\frac{2\tilde{\omega}_0}{\Delta \tilde{\omega}_0}} L, \tag{30}$$

while, for the RTM we have

$$T = \frac{2n_0}{c} L. \tag{31}$$

Using Eqs. (22), (25), (28), and (29) we find for the modulation index of the SBM:

$$\Omega_{p1} = (-1)^N \frac{\tilde{\omega}_0}{c}\sqrt{\frac{2\tilde{\omega}_0}{\Delta \tilde{\omega}_0}} \int_0^{\tilde{L}} \Delta n_{p1}(z) \cos\left(\frac{\pi N}{L} z\right) dz. \tag{32}$$

To arrive at the equivalent RTM and SBM we consider, as an example, the parametric resonance with $N = 1$ and assume that the central resonance light frequencies, resonance modulation frequencies, and modulation indices of these microresonators are equal: $\omega_0 = \tilde{\omega}_0$, $\omega_p = 2\pi/T = \tilde{\omega}_p = 2\pi/\tilde{T}$, and $\tilde{\Omega}_{p1} = \Omega_{p1}$. We calculate the integrals for the modulation indices in Eqs. (28) and (32) by setting the following similar SDM amplitudes of refractive index of the RTM and SBM (it will be shown in the next section that these distributions correspond to the SDMs with the smallest power consumption):

$$\Delta n_{p1}(z) = \Delta n_{p1}^0 \sin\left(\frac{\pi z}{L}\right), \tag{33}$$

$$\Delta \tilde{n}_{p1}(z) = \Delta \tilde{n}_{p1}^0 \cos\left(\frac{\pi z}{\tilde{L}}\right), \tag{34}$$

The spatial dependencies of these amplitudes are illustrated in Fig. 3(a). The SDM of the RTM in Eq. (33) is chosen to satisfy the noted above condition $\Delta n_{p1}(z + (L/N)) = -\Delta n_{p1}(z)$ for $N = 1$, so that it has the opposite polarity at the opposite side of the RTM as illustrated in Fig. 3(a) and (b). Generally, the SDM of an RTM can be translated along the closed RTM waveguide as a whole by substitution $z \to z - z_t$ without changing the modulation index $\Omega_{p1}$. In particular, $\sin(\dots)$ in Eq. (33) can be changed to $\cos(\dots)$. Calculation of the modulation indices using Eqs. (28), and (30)-(34) lead to the following simple relations ensuring the OFC-equivalence of the RTM and SBM:

$$\tilde{L} = g \frac{\tilde{n}_0}{n_0} L, \quad \Delta \tilde{n}_{p1}^0 = \frac{n_0}{\tilde{n}_0} \Delta n_{p1}^0, \quad g = \sqrt{\frac{2\Delta \tilde{\omega}_0}{\omega_0}}. \tag{35}$$

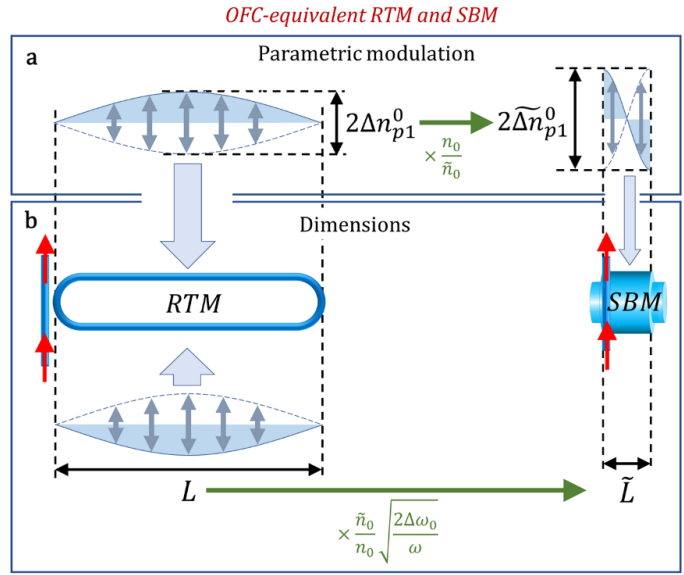

**Fig. 3.** (a) Relation between the SDMs of the refractive indices of the OFC-equivalent RTM and SBM. (b) Relation between the lengths of the OFC-equivalent RTM and SBM. In agreement with Eq. (33), the SDM along the lower part of the RTM waveguide illustrated in Fig. (b) has the opposite polarity compared to that along the upper part illustrated in Fig. (a).

The relations given by Eq. (35) between the dimensions and the amplitudes of refractive index SDMs are illustrated in Fig. 3. For the characteristic SBM bandwidth $\widetilde{\Delta \omega}_0 \simeq 2\pi \cdot 100$ GHz and light frequency $\omega_0 \simeq 2\pi \cdot 200$ THz, we find $g \simeq 0.03$, i.e., that the equivalent SBM is 30 times shorter than the RTM with the same refractive index. For the modulation frequency $\omega_p \simeq 2\pi \cdot 100$ MHz, and refractive index $n_0 \simeq 1.5$, the RTM length in this case is $L = \pi c / n_0 \omega_p \simeq 1$ m, while the length of the equivalent SBM $\tilde{L} = gL$ is around 3 mm only.

In the adiabatic case, substituting the expressions for the roundtrip times from Eqs. (30) and (31) into Eq. (27), we find the relation between the lengths of the OFC-equivalent microresonators:

$$\tilde{L} = g \frac{\Delta n_{p1}^0}{\Delta \tilde{n}_{p1}^0} L. \tag{36}$$

This relation shows that for the same lengths, $L = \tilde{L}$, the refractive index modulation amplitude, which is required to generate similar OFCs, is $g^{-1}$ times smaller for the SBM than that for the RTM. Alternatively, similar to Eq. (35), for the same amplitudes of refractive index SDMs (though not necessarily equal refractive indices), $\tilde{L}$ is $g^{-1}$ times smaller than $L$.

Calculation of the modulation index of an SBM with arbitrary shape can be reduced to the case of rectangular SBM by substitution of the axial coordinate $z$ by the time variable (Eq. (16)):

$$\tau(z,\omega) = \frac{1}{\chi} \int_0^{\tilde{z}} \frac{dz}{\beta(z,\omega)}. \tag{37}$$

Indeed, then the expression for SBM's modulation index from Eq. (22) can be written down as

$$\tilde{\Omega}_{p1} = \frac{2}{\chi} \int_0^{\tilde{T}/2} \Delta\omega_{p1}(z(\tau)) \cos\left(\tilde{\omega}_p\left(\frac{\tilde{T}}{2} - \tau\right)\right) d\tau. \tag{38}$$

Here function $z = z'(\tau)$ is the coordinate $z$ found by inversion of Eq. (37). At resonant or adiabatic condition, $\omega_p \tilde{T} = 2\pi N$, Eq. (38) is transferred to

$$\tilde{\Omega}_{p1} = (-1)^N \frac{2}{\chi} \int_0^{\tilde{T}/2} \Delta\omega_{p1}(z'(\tau)) \cos\left(\frac{2\pi\tau}{\tilde{T}}\right) d\tau. \tag{39}$$

which is similar to Eq. (32).

Small values of parameter $g$ ensure small dimensions of the SBM compared to the RTM. This advantage is achieved at the expense of the OFC bandwidth which may be limited by the bandwidth of the SBM spectrum $\widetilde{\Delta\omega}_B \sim \widetilde{\Delta\omega}_0$. To evaluate this limitation, we consider a bottle microresonator, which effective radius variation is not necessarily small, though possesses dispersionless spectrum of eigenfrequencies, as required for its effective resonant excitation. It was shown in Ref. [44] that such resonator should have the semi-cosine (rather than semi-parabolic) radius variation illustrated in Fig. 1(b),

$$\tilde{r}(z) = \begin{cases} (\tilde{r}_0 + \Delta r_0)\cos(\Delta k z) & 0 < z < \tilde{L} \\ \tilde{r}_0 & z < 0 \text{ and } z > \tilde{L} \end{cases}, \quad \Delta k = \left[(\tilde{r}_0 + \Delta r_0)R\right]^{-1/2}, \quad \tilde{L} = \frac{1}{\Delta k}\text{acos}\left(\frac{r_0}{r_0 + \Delta r_0}\right). \tag{40}$$

Such microresonator has equally spaced axial eigenfrequencies with the FSR $\widetilde{\Delta\omega}_{FSR} = 2cn_0^{-1}(\tilde{r}_0 R)^{-1/2}$. Here $R$ is the axial radius of curvature of this resonator at $z = 0$. For a shallow microresonator, $\widetilde{\Delta r_0} \ll r_0$, the profile defined by Eq. (40) becomes semi-parabolic. Generally, the microresonator height $\widetilde{\Delta r_0}$ can be much smaller as well as of the order of the radius $\tilde{r}_0$. The bandwidth of this microresonator is $\widetilde{\Delta\omega}_B = \widetilde{\Delta r_0}\tilde{\omega}_0/\tilde{r}_0$ and its length $\tilde{L}$ is defined in Eq. (40). From Eq. (31), the length of an RTM with the same eigenfrequency FSR is $L = \pi c/(n_0\widetilde{\Delta\omega}_{FSR})$. As an example, we consider, again, the resonant parametric modulation with $N = 1$ and $\tilde{\omega}_p = \widetilde{\Delta\omega}_{FSR} = 2\pi/\tilde{T}$. At this condition and for the equal resonance and modulation frequencies of the RTM and bottle microresonator, $\omega_0 = \tilde{\omega}_0$, $\omega_p = \tilde{\omega}_p$, the lengths of the RTM and SBM are:

$$L = \frac{\pi c}{n_0 \omega_p}, \quad \tilde{L} = \frac{2^{3/2} c}{\tilde{n}_0 \omega_p}\sqrt{\frac{\Delta\omega_B}{\omega_0}}. \tag{41}$$

For the maximum (to our knowledge) OFC bandwidth achieved by the parametric modulation of an RTM $\Delta\omega_B \sim$ 10 THz at $\omega_0 \sim 200$ THz [9], the OFC-equivalent SBM with the same refractive index is still 5 times shorter than the RTM. Indeed, from Eq. (41), we have $\frac{\tilde{L}}{L} = \frac{2^{\frac{3}{2}}}{\pi}\left(\frac{\Delta\omega_B}{\omega_0}\right)^{\frac{1}{2}} \sim 0.2$.

## 5. OFC optimization and power consumption

The SDM of RTMs and SBMs can be optimized to arrive at the largest possible OFC at the smallest power consumption. Here we distinguish the following cases.

(i) *Pockels modulation* when the refractive index SDM is proportional to the applied electromagnetic field. In this case, we assume that the power consumption is proportional to $P_{Pock} = \int \Delta n_{p1}(z)^2 dz$ for the RTM and to $\tilde{P}_{Pock} = \int \widetilde{\Delta n}_{p1}(z)^2 dz$ for the SBM.

(ii) *Kerr modulation* when the refractive index SDM is proportional to the absolute value of applied electromagnetic field squared. In this case, we assume that the power consumption is proportional to $P_{Kerr} = \int |\Delta n_{p1}(z)| dz$ for the RTM and to $\tilde{P}_{Kerr} = \int |\widetilde{\Delta n}_{p1}(z)| dz$ for the SBM.

While the induced parametric modulation may be extended beyond the microresonator area due to the experimental limitations, in our optimization, only the power consumed within the microresonators regions is taken into account. The method of optimization detailed in Appendix J suggests that to arrive at the strongest OFC for the given input power we have to maximize the following modulation indices, which are rescaled to fix the consumed power (see Appendix J):

$$\Omega_{p1}^{Pock} = \frac{|\Omega_{p1}|}{\sqrt{P_{Pock}}}, \quad \Omega_{p1}^{Kerr} = \frac{|\Omega_{p1}|}{P_{Kerr}}, \quad \tilde{\Omega}_{p1}^{Pock} = \frac{|\tilde{\Omega}_{p1}|}{\sqrt{\tilde{P}_{Pock}}}, \quad \tilde{\Omega}_{p1}^{Kerr} = \frac{|\tilde{\Omega}_{p1}|}{\tilde{P}_{Kerr}}. \tag{42}$$

The method of optimization detailed in Appendices J and K allows us to find the exact solutions for the optimal SDM for Pockels modulation which have to be appropriately. In contrast to the optimum Pockels modulation, which is spread along the axes of microresonators, we find below that the optimal Kerr modulation should be maximally localized at the RTM and SBM axes.

For the parametric resonant case, $\omega_p T = 2\omega_p n_0 L/c = 2\pi N$, $N = 1, 2, ...$, the optimal refractive index SDM for the Pockels modulation of an RTM is found in Appendix K as

$$\Delta n_{p1}(z) = \Delta n_{p1}^0 \cos\left(\frac{\pi N}{L}(z - z_0)\right), \tag{43}$$

where the maximum refractive index SDM amplitude $\Delta n_{p1}^0$ and shift $z_0$ are free parameters. As expected, this SDM satisfies the condition $\Delta n_{p1}(z + (L/N)) = -\Delta n_{p1}(z)$ discussed in the previous section and is translationally invariant due to the arbitrary parameter $z_0$. For the SBM at the parametric resonance, the optimal Pockels modulation of the refractive index is found as

$$\Delta n_{p1}(z) = \Delta n_{p1}^0 \frac{\beta(z_0, \omega_0)}{\beta(z, \omega_0)} \cos\left(\frac{\tilde{\omega}_p}{\chi} \int_0^z \frac{dz}{\beta(z, \omega_0)}\right) \tag{44}$$

with free parameters $\widetilde{\Delta n}_{p1}^0$ and $z_0$. For the case of rectangular SBM, the propagation constant $\beta(z, \omega_0)$ is $z$-independent and Eq. (44) coincides with Eq. (34). For the adiabatic modulation, when $N = 0$ and $\omega_p, \tilde{\omega}_p \to 0$, the optimal solutions of Eqs. (43) and (44) are simplified to $\Delta n_{p1}(z) = const$ for the RTM and $\Delta n_{p1}(z) \sim 1/\beta(z, \omega_0)$ for the SBM.

Since the accurate introduction of optimal refractive index SDMs defined by Eqs. (43) and (44) may be problematic in practice, it is important to understand the benefit of these SDMs compared to those which slightly or significantly deviate from them. In addition, it is important to evaluate the power consumption for the practically available SDMs. Below we present several examples to clarify these questions.

We start with the case of parametric resonance of an RTM with $N = 1$. For this case, the optimal refractive index SDM defined by Eq. (43) is shown in Fig. 4(a) for $z_0 = L/2$. First, we consider a partly uniform SDM, modelling the configuration of two modulated capacitors with opposite polarity considered in the experiment [9]:

$$\Delta n_{p1}(z) = \begin{cases} \Delta n_{01}^0 & z_p < z < z_p + l_p \\ -\Delta n_{01}^0 & L + z_p < z < L + z_p + l_p \\ 0 & \text{elsewhere} \end{cases} \tag{45}$$

This configuration satisfies the condition $\Delta n_{p1}(z+L) = -\Delta n_{p1}(z)$ discussed above. We find the following simple expressions for the normalized modulation indices of capacitors $\Theta_{Pock}^{(cap)}(l_p)$ and $\Theta_{Kerr}^{(cap)}(l_p)$ for the Pockels and Kerr modulations with the fixed power:

$$\Theta_{Pock}^{(cap)}(l_p) = \frac{\Omega_{p1}^{Pock}(l_p)}{\Omega_{p1}^{Pock}(L)} = \sqrt{\frac{L}{l_p}} \left| \sin\left(\frac{\pi l_p}{2L}\right) \right|, \tag{46}$$

$$\Theta_{Kerr}^{(cap)}(l_p) = \frac{\Omega_{p1}^{Kerr}(l_p)}{\Omega_{p1}^{Kerr}(L)} = \frac{L}{l_p} \left| \sin\left(\frac{\pi l_p}{2L}\right) \right|. \tag{47}$$

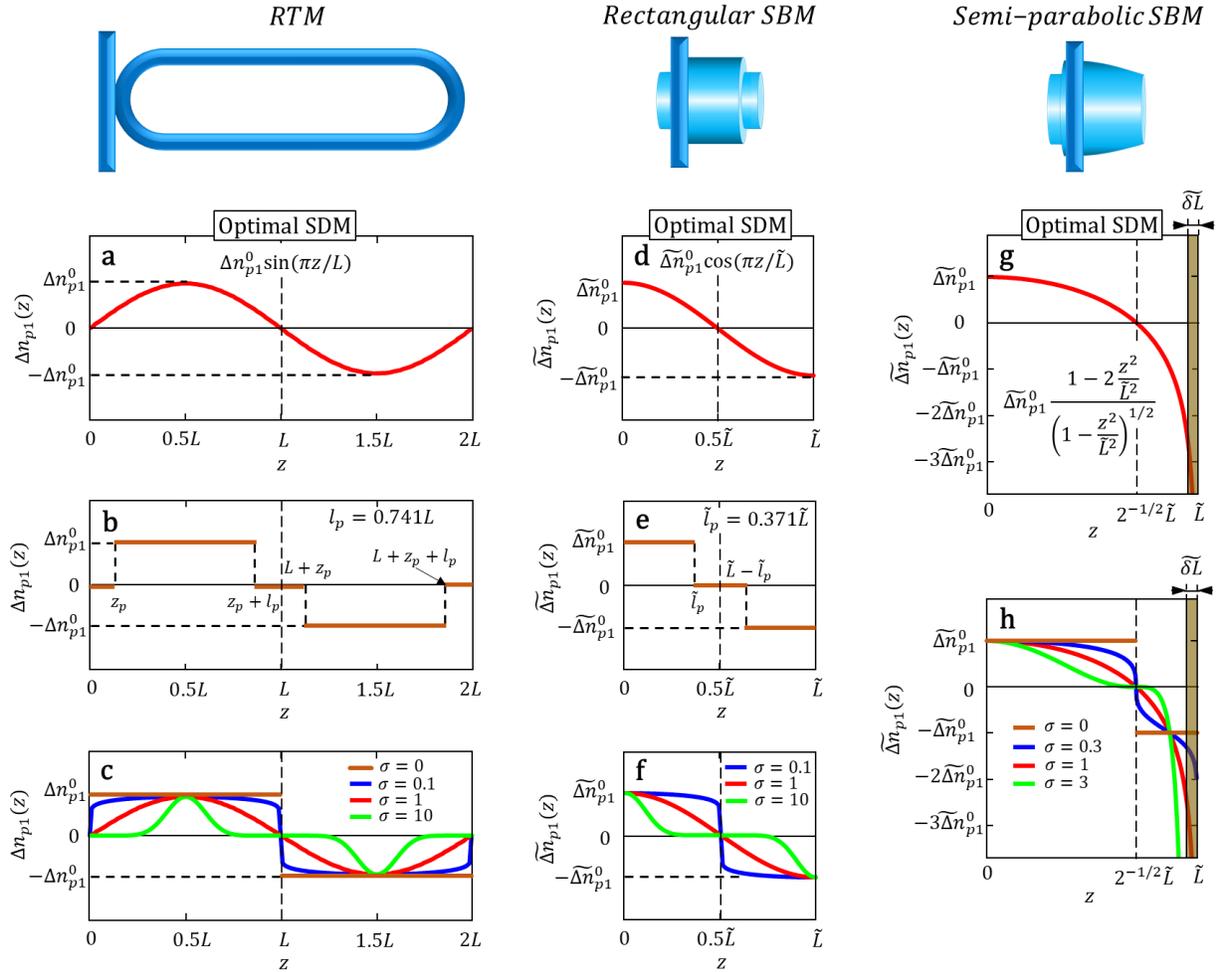

**Fig. 4.** (a) Optimal SDM of refractive index for an RTM. (b) Optimized partly uniform SDM of refractive index for an RTM. (c) SDMs of refractive index for an RTM for the model described by Eq. (48) with different values of parameter $\sigma$ indicated on the figure. (d) Optimal SDM of refractive index for a rectangular SBM. (e) Optimized partly uniform SDM of refractive index for a rectangular SBM. (f) SDMs of refractive index for a rectangular SBM for the model described by Eq. (49) with different values of parameter $\sigma$ indicated on the figure. (g) Optimal SDM of refractive index for a semi-parabolic SBM. (h) SDMs of refractive index for a semi-parabolic SBM for the model described by Eq. (53) with different values of parameter $\sigma$ indicated on the figure.

These functions determine the relative values of modulation indices found for the capacitor with length $l_p$ compared to the capacitor with length $L$ and the same power consumption. The absence of the dependence on shift $z_p$ in Eqs. (46) and (47) confirms the translational invariance of modulation. Simple calculations show that the maximum of $\Theta_{Pock}^{(cap)}(l_p)$ is achieved at $l_p = 0.741L$ (Fig. 4(b)) for the Pockels modulation and at $l_p \to 0$ for the Kerr modulation. The dependencies $\Theta_{Pock}^{(cap)}(l_p)$ and $\Theta_{Kerr}^{(cap)}(l_p)$ are shown in Fig. 5(a). For the Pockels modulation, the power benefit of the optimized capacitor length $l_p = 0.741L$ compared to the length $l_p = L$ is 7% only. For the Kerr modulation, this power consumption can be $\frac{\pi}{2} \cong 1.57$ times smaller for $l_p \to 0$ compared to $l_p = L$.

Consider now a rectangular SBM with a partly uniform refractive index SDM similar to that described by Eq. (45), where now we have to replace $L \to \tilde{L}/2$, $z_p \to 0$, $l_p \to \tilde{l}_p$, $\Delta n_{p1}(z) \to \widetilde{\Delta n}_{p1}(z)$, and $\Delta n_{p1}^0 \to \widetilde{\Delta n}_{p1}^0$. The optimum SDM in this case following from Eq. (44) is shown in Fig. 4(d). Calculations show that, for the Pockels modulation, the smallest power consumption is achieved at $\tilde{l}_p = 0.371L$ (Fig. 4(e)) with the 7% benefit in power consumption compared to $\tilde{l}_p = 0.5\tilde{L}$. For the Kerr modulation, the optimum $\tilde{l}_p \to 0$ and the power consumption can be $\frac{\pi}{2}$ times smaller, similar to the RTM.

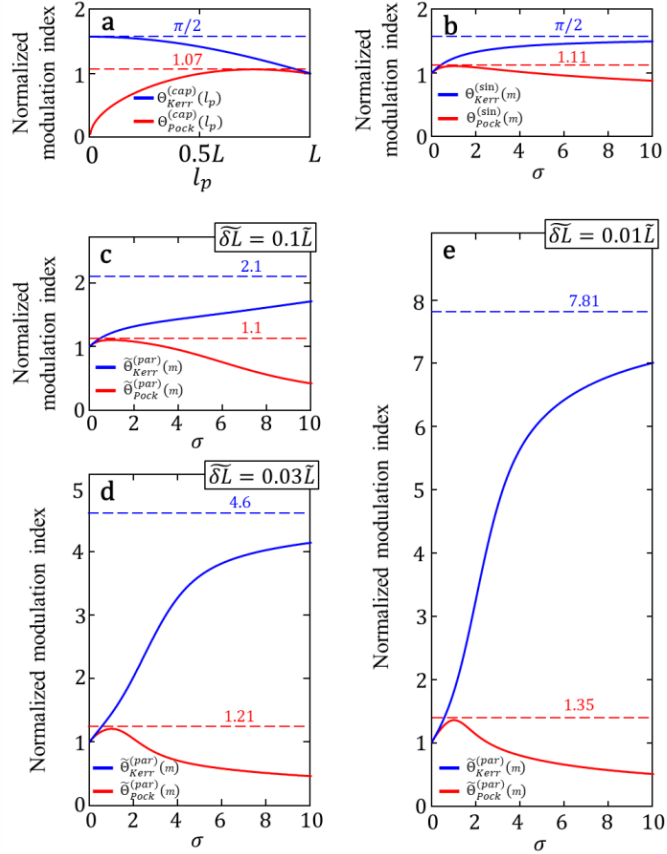

**Fig. 5.** Normalized modulation indices as functions of the refractive index SDM parameters for the Pockels modulation (red curves) and Kerr modulation (blue curves): (a) for a partly uniform SDM, which models two capacitors with opposite polarity adjacent to an RTM as a function of the capacitor length $l_p$; (b) for the SDM of an RTM described by Eq. (48) as a function of parameter $\sigma$; (c), (d), and (e) for the SDM of a semi-parabolic SBM described by Eq. (53) as a function of parameter $\sigma$ calculated for different integration cutoff lengths $\widetilde{\Delta L}$ indicated on the figures.

The spatially uniform SDM considered above can be induced by capacitors adjacent to the microresonators. In a more general case, the refractive index SDM is nonuniformly distributed along the microresonator length.

Such SDM can be induced by electromagnetic field of nonuniform capacitors as well as other types of electromagnetic field sources. As an example, we consider the following SDMs of refractive index of an RTM and a rectangular SBM (Fig. 4(c) and (f)):

$$\Delta n_{p1}(z) = n_{p1}^0 \left|\sin\left(\frac{\pi z}{L}\right)\right|^\sigma \text{sign}\left(\sin\left(\frac{\pi z}{L}\right)\right), \tag{48}$$

$$\Delta n_{p1}(z) = \tilde{n}_{p1}^0 \left|\cos\left(\frac{\pi z}{\tilde{L}}\right)\right|^\sigma \text{sign}\left(\cos\left(\frac{\pi z}{\tilde{L}}\right)\right), \tag{49}$$

with parameter $\sigma$. As previously, we consider the case of parametric resonance with $N = 1$. At $\sigma \to 0$, these SDMs coincide with the previous examples at $l_p = L$ for the RTM (compare Figs. 4(b) and (c)) and $\tilde{l}_p = 0.5\tilde{L}$ for the SBM (compare Figs. 4(e) and (f)). At $\sigma = 1$, these SDMs correspond to the minimum power consumption (see Eq. (43), (44), and (34)).

We calculate the rescaled modulation indices defined by Eq. (42) as a function of parameter $\sigma$ and introduce normalized modulation indices $\Theta_{Pock}^{(\sin)}(\sigma)$, $\Theta_{Kerr}^{(\sin)}(\sigma)$, $\tilde{\Theta}_{Pock}^{(\sin)}(\sigma)$, and $\tilde{\Theta}_{Kerr}^{(\sin)}(\sigma)$ for the Pockels and Kerr modulations with SDMs defined by Eqs. (48) and (49) similar to those in Eqs. (46) and (47):

$$\Theta_{Pock}^{(\sin)}(\sigma) = \frac{\Omega_{p1}^{Pock}(\sigma)}{\Omega_{p1}^{Pock}(0)}, \quad \tilde{\Theta}_{Pock}^{(\sin)}(\sigma) = \frac{\tilde{\Omega}_{p1}^{Pock}(\sigma)}{\tilde{\Omega}_{p1}^{Pock}(0)}, \quad \Theta_{Kerr}^{(\sin)}(\sigma) = \frac{\Omega_{p1}^{Kerr}(\sigma)}{\Omega_{p1}^{Kerr}(0)}, \quad \tilde{\Theta}_{Kerr}^{(\sin)}(\sigma) = \frac{\tilde{\Omega}_{p1}^{Kerr}(\sigma)}{\tilde{\Omega}_{p1}^{Kerr}(0)}. \tag{50}$$

These functions determine the ratio of modulation indices with parameters $\sigma$ and $\sigma = 0$ at the same modulation power. For the Pockels modulation of an RTM, the plots of $\Theta_{Pock}^{(\sin)}(\sigma)$ and $\Theta_{Kerr}^{(\sin)}(\sigma)$ are shown in Fig. 5(b). It is seen that, as expected, the power consumption achieves minimum at $m = 1$ with the SDM shown by the red curve in Figs. 4(a) and (c). The power benefit of the optimized modulation compared to the case $l_p = 0.5L$ considered above is 11%. Thus, the application of optimized nonuniform capacitors can slightly improve the performance of the system considered in Ref. [9]. For the Kerr modulation of an RTM, the power consumption achieves minimum at $\sigma \to \infty$ (Fig. 5(b), blue curve), i.e., when the SDM is strongly localized. Then, the consumed power for the optimized SDM is $\frac{\pi}{2}$ times smaller compared to the case $l_p = L$ as in the case of partly uniform SDM considered above.

The results for the SBM modulation are similar. For the Pockels modulation, the optimum refractive index SDM corresponds to $\sigma = 1$ (red curve in Figs. 4(d) and (f)) with 11% power benefit compared to the case $\tilde{l}_p = 0.5\tilde{L}$ considered above. For the Kerr modulation, the power consumption is, again, minimised at $\sigma \to \infty$ and is $\frac{\pi}{2}$ times smaller compared to the case $\tilde{l}_p = 0.5\tilde{L}$.

Finally, assume that an SBM has a semi-parabolic shape, with effective radius variation $r(z) = r_0 + \widetilde{\Delta r_0} - z^2/2R$ and length $\tilde{L} = (2\widetilde{\Delta r_0}R)^{1/2}$, so that its axial eigenfrequency spectrum is dispersionless having a constant FSR $\widetilde{\Delta \omega}_{FSR} = 2cn_0^{-1}(\tilde{r}_0 R)^{-1/2}$ (compare with Eq. (40)). Then, at the parametric resonant condition, $\omega_p = \widetilde{\Delta \omega}_{FSR} = 2\pi/\tilde{T}$, Eq. (22) for the modulation index and optimum refractive index SDM (Eq. (43)) are simplified to

$$\tilde{\Omega}_{p1} = -\frac{2\omega_0 (Rr_0)^{1/2}}{c\tilde{L}} \int_0^{\tilde{L}} \Delta n_{p1}(z) G(z) dz, \quad G(z) = \left(1 - 2\frac{z^2}{\tilde{L}^2}\right)\left(1 - \frac{z^2}{\tilde{L}^2}\right)^{-1/2}, \tag{51}$$

$$\Delta n_{p1}(z) = \Delta n_{p1}^0 G(z). \tag{52}$$

In derivation of these expressions, we used the rescaling relation, Eq. (26). In contrast to the rectangular SBM, with the sine-shaped kernel in the integral for $\widetilde{\Omega}_{p1}$ (Eq. (32)), the kernel $G(z)$ of the integral in Eq. (51) and the optimal $\widetilde{\Delta n}_{p1}(z)$ (shown in Fig. 4(g)) are aperiodic and have a singularity at the turning point $z = \tilde{L}$. In the neighbourhood of this turning point, the WGM propagation speed tends to zero and the semiclassical approximation used here requires corrections unless the contribution from this neighbourhood into the integral in Eq. (51) is small.

To investigate the effect of deviation from the optimal $\widetilde{\Delta n}_{p1}(z)$ defined by Eq. (52), we consider the refractive index SDM

$$\Delta n_{p1}(z) = \Delta n_{p1}^{0} \left(G(z)\right)^{\sigma} \text{sign}\left(G(z)\right). \tag{53}$$

depending on parameter $\sigma$ and illustrated in Fig. 4(h) for $\sigma = 0, 0.3, 1,$ and $3$. This SDM is partly uniform at $\sigma \to 0$, optimal at $\sigma = 1$ (when $\Delta n_{p1}(z) \sim G(z)$), and strongly localized at $\sigma \to \infty$. Similar to Eq. (50), we introduce the ratios of modulation indices with parameters $\sigma$ and $\sigma = 0$ at the same modulation power:

$$\Theta_{Pock}^{(par)}(\sigma) = \frac{\Omega_{p1}^{Pock}(\sigma)}{\Omega_{p1}^{Pock}(0)}, \quad \widetilde{\Theta}_{Pock}^{(par)}(\sigma) = \frac{\widetilde{\Omega}_{p1}^{Pock}(\sigma)}{\widetilde{\Omega}_{p1}^{Pock}(0)}, \quad \Theta_{Kerr}^{(par)}(\sigma) = \frac{\Omega_{p1}^{Kerr}(\sigma)}{\Omega_{p1}^{Kerr}(0)}, \quad \widetilde{\Theta}_{Kerr}^{(par)}(\sigma) = \frac{\widetilde{\Omega}_{p1}^{Kerr}(\sigma)}{\widetilde{\Omega}_{p1}^{Kerr}(0)}. \tag{54}$$

We estimate the size of vicinity near $z = \widetilde{L}$ where the expressions in Eqs. (51) and (52) fail as having the order of $\widetilde{\delta L} = (c^2 r_0 R)^{\frac{1}{3}} (\omega_0^2 n_0^2 \widetilde{L})^{-1/3}$. This expression is found from the asymptotic of an eigenfunction of the parabolic SBM [45] near the turning point $z = \widetilde{L}$. Since the integral for the modulation index in Eq. (51) calculated with the refractive index SDM of Eq. (53) diverges at $\sigma > \frac{1}{2}$, in our calculations, we cut the integral limit in Eq. (51) by replacing $\widetilde{L} \to \widetilde{L} - \widetilde{\delta L}$. We estimate the cutoff region $\widetilde{\delta L}$, illustrated as a brown area in Figs. 4(g) and (h), for the parameters of semi-parabolic SBM experimentally investigated in Ref. [24] setting $L = 3$ mm, $r_0 = 20$ μm, $R = 120$ m, $n_0 = 1.46$, and $\omega_0 = 2\pi \cdot 190$ THz, which yields $\widetilde{\delta L} = 90$ μm $\sim 0.03L$. Consequently, Figs. 5(c), (d), and (e) show the plots of normalized modulation indices defined by Eq. (54) which are calculated for $\widetilde{\delta L} = 0.1L, 0.03L$ and $0.01L$. It is seen that, for the Pockels modulation, the optimal SDM only weakly depends on $\widetilde{\delta L}$. However, the power consumption for the Kerr modulation can be significantly reduced with $\widetilde{\delta L}$.

## 6. Discussion

This work is primary stimulated by recent experimental results and theoretical analysis of OFC generation by parametric modulation of microresonators [9, 10] where the OFCs were generated by an RTM [9] and a disk microresonator [10] both fabricated of Lithium Niobate. To describe the OFC generation, the authors of Ref. [10] further developed the model Hamiltonian approach which was pioneered for the parametric modulation of optical microresonators in Ref. [14]. This theory assumes that the modulation index is small and only single quantum transitions are possible during a single circulation of light around the microresonator. For the same purpose, the authors of Ref. [9] used a semi-empirical theory developed in Ref. [12] to describe the OFC generation by parametric modulation of an RTM. This theory does not restrict the value of modulation index so that it can be smaller as well as greater than unity. Therefore, the multiquantum transitions of light during its single roundtrip along the RTM are taken into account. For small modulation indices, the theories [12, 9] and [14, 10] give similar results. In both theories the modulation index was introduced as a real parameter which relation to the spatial distribution of modulation (SDM) of the microresonator parameters has not been investigated.

Here, we develop the semiclassical perturbation theory which allows us to understand, describe, and quantify the key effects in multi-quantum parametric modulation of optical microresonators and OFC generation. In comparison to the theories developed previously (see [9, 10] and references therein), our approach allows us to directly express the basic parameters describing OFC generation, such as modulation index, through the SDM of microresonator parameters (e.g., refractive index). In particular, these results allow us to optimize the SDM along the microresonator length.

Targeting at a more transparent and focused presentation of the material, we have moved several important results of the paper, including explanations of calculations made, into the Appendices. We note in the main text that the solution of the major equations used to derive the expressions for the OFC amplitudes are reduced to functional equations. These equations are solved in Appendices B and C. In the main text of the paper, we consider the generation of OFC by RTMs and SBMs coupled to the input-output waveguide (Fig. 1). In Sections 2 and 3, we derive analytical expressions for the OFC amplitudes generated in RTMs and SBMs. In contrast to the previous results, simple analytical formulas for the complex-valued modulation indices expressed through the spatially distributed parametric (refractive index for the RTM and WGM cutoff frequency for the SBM) modulations are presented. Our results allow to understand and quantify the characteristic features of OFC in both types of microresonators, e.g., the resonance splitting effect (Fig. 2). The results of these sections can be better understood after the analysis of the behaviour of parametrically modulated *free* microresonators considered in Appendix D for the RTM and in Appendix G for the SBM. Complementary to Sections 2 and 3, in Appendix M, we consider the OFC generation in modulated microresonators excited by an internal light source, which is used in many theoretical studies of microresonators (see e.g., [10, 31, 46] and references therein), and find the direct relation between this model and the model of the input-output waveguide used in practice. For the purpose of a more straightforward comparison of the performance of the RTM and SBM, in the main text of this paper we consider the input-output waveguides positioned at the edge of SBMs. In Appendix L we describe the approach, which allows to generalise our results to the case of the arbitrary waveguide position.

In Section 4 we compare the performance of RTMs and SBMs. We introduce the OFC-equivalent RTM and SBM as microresonators which generate identical OFCs and find simple criterium of this equivalence. In an SBM, light is circulating along the microresonator circumference in the process of slow propagation along its axis. For this reason, light in an SBM can cover much longer distance during one roundtrip along its axis compared to the distance the light propagates covering a roundtrip along the RTM waveguide. For this reason, the dimensions of an SBM which is OFC-equivalent to an RTM can be much smaller than those of the RTM (Fig. 3). We note in Section 4 that the broadband OFC can be generated using a bottle microresonator with effective radius variation which is not necessarily small as required in the SNAP theory. As shown in Ref. [44], such resonators can still have the dispersionless distribution of axial eigenfrequencies. Another possibility to expand the OFC spectral band consists in matching the axial and azimuthal eigenfrequencies, as suggested in Ref. [47] for the nonlinear OFC generation.

In Section 5, we optimize the spatial distribution of parametric modulation for RTMs and SBMs to arrive at the smallest consumption power. We distinguish the Pockels modulation, when this power is proportional to the amplitude of refractive index modulation, and the Kerr modulation, when it is proportional to this amplitude squared. As shown in Appendix K, for the Pockels modulation, the developed theory allows to find the optimal spatial distributions of modulations analytically both for the RTM and SBM. While, for the Pockels modulation, these distributions are spread along the lengths of microresonators, the optimal Kerr modulation is localized. In practice, the optimal modulation distributions found in this Section are not easy to realize. For this reason, we compare them with other distribution models (Figs. 4 and 5). Generally, the developed theory allows to determine and optimize the modulation distribution with constrains determined by their practical availability.

**Acknowledgements**

The work on the paper was supported by the Engineering and Physical Sciences Research Council (EPSRC) (grants EP/P006183/1 and EP/W002868/1) and Leverhulme Trust (grant RPG-2022-014).

**Appendix A. Solution of the first order partial differential equation describing semiclassical propagation of light along the modulated waveguide**

We are looking for the solution of the wave equation, Eq. (1) of the main text, with a source $I_{in}(x,t)$,

$$\frac{\partial^2 E}{\partial z^2} - \frac{1}{c^2}\frac{\partial^2 D}{\partial t^2} = I_{in}(z,t), \tag{A1}$$

under the condition that the source $I_{in}(x,t)$ has frequency $\omega$ and spatial distribution which smoothly envelops the wave with propagation constant $\beta$:

$$I_{in}(z,t) = \exp(-i\omega t + i\beta z) F_{in}(z), \quad \beta = \frac{\omega n_0}{c}. \tag{A2}$$

Here $F_{in}(x)$ varies slowly compared to $\exp(ikx)$. The semiclassical solution of Eq. (A1) is found as

$$E(z,t) = \exp(-i\omega t + \beta z)\Psi(z,t) \tag{A3}$$

where $\Psi(x,t)$ satisfies the first order partial differential equation

$$\frac{\partial \Psi}{\partial t} + \frac{c}{n_0}\frac{\partial \Psi}{\partial z} - \frac{i\omega}{n_0}\Delta n(z,t)\Psi = \frac{\exp(-i\Delta\omega t)}{2i\omega_0}F_{in}(z), \quad \Delta n(z,t) = n(z,t) - n_0. \tag{A4}$$

Solution of Eq. (A4) can be found analytically in the form [48]:

$$\Psi(z,t) = U(z,t)\left\{\Phi\left(t - \frac{n_0}{c}z\right) + \frac{n_0}{2ic\omega}\int_0^z dz' \frac{F_{in}(z')}{U(z',t)}\right\},$$

$$U(z,t) = \exp\left[i\frac{\omega}{c}\int_0^z dz'\Delta n\left(z', t - \frac{n_0}{c}(z-z')\right)\right], \tag{A5}$$

with arbitrary function $\Phi(t)$. In the absence of source $I_{in}(x,t)$, Eq. (A5) is simplified to

$$\Psi(z,t) = U(z,t)\Phi\left(t - \frac{n_0}{c}z\right),$$

$$U(z,t) = \exp\left[i\frac{\omega}{c}\int_0^z dz'\Delta n\left(z', t - \frac{n_0}{c}(z-z')\right)\right]. \tag{A6}$$

**Appendix B. Solution of the functional equation $\Phi(t) = A(t)\Phi(t-T) + B(t)$**

We solve equation

$$\Phi(t) = A(t)\Phi(t-T) + B(t) \tag{B1}$$

by substitution

$$\Phi(t) = C(t)\Phi_1(t) \tag{B2}$$

which yields

$$\Phi_1(t) = \frac{A(t)C(t-T)}{C(t)}\Phi_1(t-T) + \frac{B(t)}{C(t)} \tag{B3}$$

We determine function $C(t)$ so that the coefficient in front of $\Phi_1(t-T)$ in Eq. (B3) is equal to unity:

$$C(t) = A(t)C(t-T) \tag{B4}$$

It is easy to verify by direct substitution that one of solutions of this equation is

$$C(t) = D_1(t)\prod_{n=0}^{\infty} A(t-nT) \tag{B5}$$

Here $D_1(t)$ is an arbitrary periodic function with period $T$. Thus, Eq. (B3) takes the form:

$$\Phi_1(t) = \Phi_1(t-T) + \frac{B(t)}{C(t)} \tag{B6}$$

Solution of this equation is:

$$\Phi_1(t) = \sum_{n=0}^{\infty} \frac{B(t-nT)}{C(t-nT)} + D_2(t) \tag{B7}$$

where $D_2(t)$ is an arbitrary periodic function with period $T$. Note that solution of Eq. (B6) can be found from the solution of Eq. (B4) by taking the logarithm of both parts of Eq. (B4).

Combining Eqs. (B2), (B5) and (B7) we find:

$$\Phi(t) = B(t) + \sum_{n=1}^{\infty} B(t-nT) \prod_{m=0}^{n-1} A(t-mT) + D(t) \prod_{m=0}^{\infty} A(t-mT) \tag{B8}$$

where $D(t)$ is an arbitrary periodic function with period $T$.

The general solution of Eq. (B1) can be found by adding the general solution of the uniform equation

$$\Phi_0(t) = A(t)\Phi_0(t-T) \tag{B9}$$

to the particular solution $\Phi(t)$ of Eq. (B1) defined by Eq. (B8). Since function $A(t)$ of our concern has the period $T_p = \frac{2\pi}{\omega_p}$ and, therefore, can be presented in the form

$$A(t) = \exp(ia(t)), \quad a(t) = \sum a_m \exp(im\omega_p t), \tag{B10}$$

we are looking for solution of Eq. (B9) in the form of the Floquet solution

$$\Phi_0(t) = D_0(t)\exp(i\alpha t + if(t)), \quad f(t) = \sum_{\substack{m=-\infty \\ m\neq 0}}^{\infty} f_m \exp(im\omega_p t), \tag{B11}$$

where $D_0(t)$ is an arbitrary $T$-periodic function of $t$. Eqs. (B9)-(B11) yield:

$$\alpha = \frac{a_0 + 2\pi q}{T}, \quad q = 0, \pm 1, \pm 2, \ldots$$

$$f_m = \frac{a_m}{1 - \exp(-i\omega_p T)} \quad \text{for } m \neq 0. \tag{B12}$$

Finally, solution of Eq. (B1) is a sum of solution (B8) and (B11) with parameters defined by Eq. (B12).

**Appendix C. Solution of the functional equation for the field in a RTM coupled to a waveguide**

From Eq. (3) of the main text, we find

$$E_{out}(t) = \tau E_{in}(t) + \kappa E(2L,t) \tag{C1}$$

$$E(0,t) = -\kappa E_{in}(t) + \tau E(2L,t) \tag{C2}$$

Substitution of the input field $E_{in}(t) = \exp(-i\omega t)$ and functions $E(0,t)$ and $E(2L,t)$ from Eq. (D2) into Eq. (C2) yields the functional equation for $\Phi(t)$:

$$\Phi(t) = \tau A(t)\Phi(t-T) - \kappa,$$

$$A(t) = \exp\left[i\omega T - \frac{2\eta\omega L}{c} + i\Omega_{p0} + i|\Omega_{p1}|\cos\left(\omega_p(t-T) + \arg(\Omega_{p1})\right)\right] \quad (C3)$$

This equation is a particular case of Eq. (B1). Taking into account that, for finite $\eta$, $\prod_{m=0}^{\infty} A(t-mT) = 0$ and that the solution of our concern tends to zero for $\kappa \to 0$, we find from Eqs. (B8) and (C3):

$$\Phi(t) = -\kappa - \kappa \sum_{n=1}^{\infty} \tau^n \prod_{m=0}^{n-1} A(t-mT) \quad (C4)$$

Combining Eqs. (C1), (D2), and (C4) we find:

$$\begin{aligned}
E_{out}(t) &= E_{in}(t)\tau\left(1 + \kappa A(t)\Phi(t-T)\right) \\
&= E_{in}(t)\tau\left\{1 - \kappa^2 A(t)\left[1 + \sum_{n=1}^{\infty} \tau^n \prod_{m=0}^{n-1} A(t-(m+1)T)\right]\right\} \\
&= E_{in}(t)\tau\left\{1 - \kappa^2 A(t)\left[1 + \sum_{n=1}^{\infty} \tau^n \prod_{m=1}^{n} A(t-mT)\right]\right\} \\
&= E_{in}(t)\tau\left\{1 - \kappa^2 \sum_{n=0}^{\infty} \tau^n \prod_{m=0}^{n} A(t-mT)\right\}
\end{aligned} \quad (C5)$$

Next, we calculate

$$\begin{aligned}
\prod_{m=0}^{n} A(t-mT) &= \exp\left[(n+1)\left(i\omega T - \frac{2\eta\omega L}{c} + i\Omega_{p0}\right)\right] \exp\left[i|\Omega_{p1}| \sum_{m=0}^{n} \cos\left(\omega_p(t-T-mT) + \arg(\Omega_{p1})\right)\right] \\
&= \exp\left[(n+1)\left(i\omega T - \frac{2\eta\omega L}{c} + i\Omega_{p0}\right)\right] \exp\left[i\sigma_{n+1} \cos\left(\omega_p t - \frac{(n+2)}{2}\omega_p T + \arg(\Omega_{p1})\right)\right],
\end{aligned} \quad (C6)$$

$$\sigma_n = |\Omega_{p1}| \frac{\sin\left(\frac{n}{2}\omega_p T\right)}{\sin\left(\frac{1}{2}\omega_p T\right)}.$$

In the derivation of the latter expression, we used the identity [49]:

$$\sum_{k=0}^{n} \cos(kx + a) = \frac{\sin\left(\frac{(n+1)x}{2}\right)}{\sin\left(\frac{x}{2}\right)} \cos\left(\frac{nx}{2} + a\right) \quad (C7)$$

Combining Eqs. (C5) and (C6), we find:

$$E_{out}(t) = E_{in}(t)\tau\left\{1 - \kappa^2 \sum_{n=0}^{\infty} \tau^n \exp\left[(n+1)\left(i\omega T - \frac{2\eta\omega L}{c} + i\Omega_{p0}\right)\right. \right.$$
$$\left.\left. + i\sigma_{n+1} \cos\left(\omega_p t - \frac{n+2}{2}\omega_p T + \arg(\Omega_{p1})\right)\right]\right\}. \quad (C8)$$

Application of the Jacobi-Anger formula to $\exp(i\sigma_{n+1}\cos(...))$ in Eq. (C8) allows to determine the comb spectral components $E_m$ of $E_{out}(t)$ given by Eq. (4) of the main text.

Close to the condition of parametric resonance or adiabatic modulation, $\omega_p T = 2\pi N + \varepsilon$, $N = 0, 1, 2, ..., \varepsilon \ll 1$, we have

$$\sigma_{n+1} = (-1)^{Nn}(n+1)|\Omega_{p1}|, \tag{C10}$$

so that in Eq. (C8)

$$\sigma_{n+1}\cos\left(\omega_p t - \frac{n+2}{2}\omega_p T + \arg(\Omega_{p1})\right) = (-1)^{Nn}(n+1)|\Omega_{p1}|\cos(\omega_p t - \pi Nn + \arg(\Omega_{p1}))$$
$$= (n+1)|\Omega_{p1}|\cos(\omega_p t + \arg(\Omega_{p1})) \tag{C11}$$

Then, the sum in Eq. (C8) becomes a geometric series easy to calculate. In this case, function $A(t)$ is a $T$-periodic function, $A(t-T) = A(t)$ and we have from Eq. (C5):

$$E_{out}(0,t) = E_{in}(2L,t)\tau\left(1 - \kappa^2 \sum_{n=0}^{\infty} \tau^n \prod_{m=0}^{n} A(t-mT)\right) = E_{in}(2L,t)\tau\left(1 - \kappa^2 \sum_{n=0}^{\infty} \tau^n \prod_{m=0}^{n} A(t)\right)$$
$$= E_{in}(2L,t)\tau\left(1 - \kappa^2 A(t)\sum_{n=0}^{\infty} \tau^n A(t)^n\right) = E_{in}(2L,t)\tau\left(1 - \kappa^2 \frac{A(t)}{1-\tau A(t)}\right) \tag{C12}$$
$$= E_{in}(2L,t)\frac{\tau - A(t)}{1-\tau A(t)}$$

where the identity $\kappa^2 + \tau^2 = 1$ was used. This result coincides with Eq. (9) after introducing the resonance frequency $\omega_q$ satisfying Eq. (10).

**Appendix D. Quasi-states of a standing along (uncoupled) RTM**

From Eq. (A5), in the absence of source, $F_{in}(x) = 0$, the general solution of Eqs. (1) and (A1) is

$$E(z,t) = \exp\left[-i\omega t + i\beta z + i\frac{\omega}{c}\int_0^z dz' \Delta n\left(z', t - \frac{n_0}{c}(z-z')\right)\right]\Phi\left(t - \frac{n_0}{c}z\right), \tag{D1}$$

with $\beta = \frac{\omega n_0}{c}$ and $\Delta n(z,t) = n(z,t) - n_0$ where $n(z,t)$ is defined by Eq. (2). We find from Eq. (D1):

$$E(0,t) = \exp(-i\omega t)\Phi(t),$$
$$E(2L,t) = \exp(-i\omega t)A(t)\Phi(t-T),$$
$$A(t) = \exp\left[i\omega T + i\Omega_{p0} - \frac{2\eta\omega L}{c} + i|\Omega_{p1}|\cos(\omega_p(t-T) + \arg(\Omega_{p1}))\right] \tag{D2}$$
$$T = \frac{2n_0 L}{c}.$$

where parameters $\Omega_{p0}$ and $\Omega_{p1}$ are defined by Eq. (6). For a free RTM, this solution and its spatial derivative should be continuous along the resonator length, i.e., satisfy the conditions: $E(0,t) = E(2L,t)$ and $\frac{dE}{dz}(0,t) = \frac{dE}{dz}(2L,t)$. In the semiclassical approximation considered, function $\Phi$ is a much slower function of $z$ than $\exp(i\beta z)$ in Eq. (D1). Therefore, these two conditions coincide and lead to the functional equation for $\Phi(t)$:

$$\Phi(t) = A(t)\Phi(t-T), \tag{D3}$$

We introduce the quasi-eigenfrequency $\omega_0$ by the quantization rule:

$$\omega_0 T + \Omega_{p0} = 2\pi n_0, \quad n_0 \gg 1, \text{ integer.} \tag{D4}$$

Solution of Eq. (D3) is found in Appendix B. For $A(t)$ defined by Eqs. (D2) with frequency $\omega = \omega_0$, parameters in Eq. (B12) are:

$$a_0 = \frac{2\eta\omega_0 L}{c}i, \quad a_1 = \frac{\Omega_{p1}}{2}, \quad a_{-1} = a_1^*, \quad a_m = 0 \text{ for } |m| > 1. \tag{D5}$$

Solutions given by Eqs. (D1), (D4), (B12) and (D5), allow us to find the general expressions for the quasi-eigenfrequencies and quasi-states. Here, for briefness and without loss of generality, we assume that $\Delta n(z,t) = 0$ in the neighborhood of $z = 0$ in the region $0 < z < z_p$ and $2L - z_p < z < 2L$ shown in Fig. 1(a). In this region we find for the normalized at $t = 0$ quasi-states:

$$E_q(z,t) = (2L)^{-1/2} \exp\left[-i\omega_0\left(t - \frac{n_0 z}{c}\right)\right] \Psi_q(z,t),$$

$$\Psi_q(z,t) = \exp\left[-i\frac{2\pi q}{T}\left(t - \frac{n_0 z}{c}\right) - \frac{\eta}{n_0}\omega_0 t + i|\Upsilon_p|\cos\left(\omega_p\left(t - \frac{n_0 z}{c}\right) + \arg(\Upsilon_p)\right)\right], \tag{D6}$$

The eigenfrequencies of these quasi-states are determined as

$$\omega_q = \omega_0 + \frac{2\pi q}{T}, \quad q = 0, \pm 1, \pm 2, ..., \quad \omega_0 = \frac{2\pi s - \Omega_{p0}}{T}, \quad |q| \ll s. \tag{D7}$$

The integer $s \gg 1$ in Eq. (D7) is chosen so that the eigenfrequency $\omega_0$ is close to the frequency of our interest and the *quasi-state modulation index* $\Upsilon_p$ in Eq. (D6) is expressed through the modulation index $\Omega_{p1}$ as

$$\Upsilon_p = \frac{\Omega_{p1}}{\exp(i\omega_p T) - 1}. \tag{D8}$$

Similar expressions for the quasi-states can be obtained if we introduce complex-values quasi-eigenfrequencies by substitution $\omega_q \to \omega_q - i\eta\omega_0/n_0$ and remove the attenuating factor $\exp[-(\eta\omega_0/n_0)t]$ from Eq. (D6). In this equation, we ignored the attenuation along the RTM axis $z$ due to the relatively large Q-factor of RTM leading to $\frac{\omega_0 \eta T}{n_0} \ll 1$. However, the latter attenuation contributes to the widths of transmission resonances and cannot be ignored in the consideration of an RTM coupled to the input-output waveguide.

Quasi-states $E_q(z,t)$ in Eq. (D6) are presented as a product of fast oscillating exponent $\exp(-i\omega_0 t + i\beta_0 z)$ and relatively slow function of time and space $\Psi_q(x,t)$ which contains the key information about the quasi-state behaviour. For the microresonators of our interest, the intrinsic quasi-state Q-factor, $Q = \frac{n_0}{2\eta}$, is large, $Q \gg 1$ (typically, $Q \sim 10^5 - 10^{10}$). The attenuation of $E_q(x,t)$ in time assumed here is relatively slow, $\frac{\omega_0 T \eta}{n_0} = \frac{\omega_0 T}{2Q} \ll 1$. The latter condition follows from $Q \gg 1$ for the cases we are primary interested in this paper: the resonant modulation, $\omega_0 T \cong 2\pi N$, $N = 1, 2, 3, ...$, and adiabatic modulation, $\omega_0 T \ll 1$.

Slow function $\Psi_q(x,t)$ in Eq. (D6) is a product of two periodic functions with periods $T = \frac{2n_0 L}{c}$ and $T_p = \frac{2\pi}{\omega_p}$, where $T$ determines the time of circulation of light along the full microresonator length $2L$ and $T_p$ is the parametric modulation period. As follows from Eq. (D6), the effect of modulation is characterized by the quasi-state modulation index $\Upsilon_p$ determined by Eq. (D8). In order to find the frequency comb spectral components of quasi-states $E_q(z,t)$, we expand it into series of harmonics using the Jacobi-Anger formula:

$$E_q(z,t) = \sum_{m=-\infty}^{\infty} E_{0m} \exp\left\{-i\left[\omega_0 + \frac{2\pi q}{T} - \omega_p m\right]\left(t - \frac{n_0 z}{c}\right) - \frac{\eta}{n_0}\omega_0 t\right\},$$

$$E_{0m} = (2L)^{-1/2} \exp\left[im\left(\frac{\pi}{2} + \arg(\Upsilon_p)\right)\right] J_m(|\Upsilon_p|).$$

(D9)

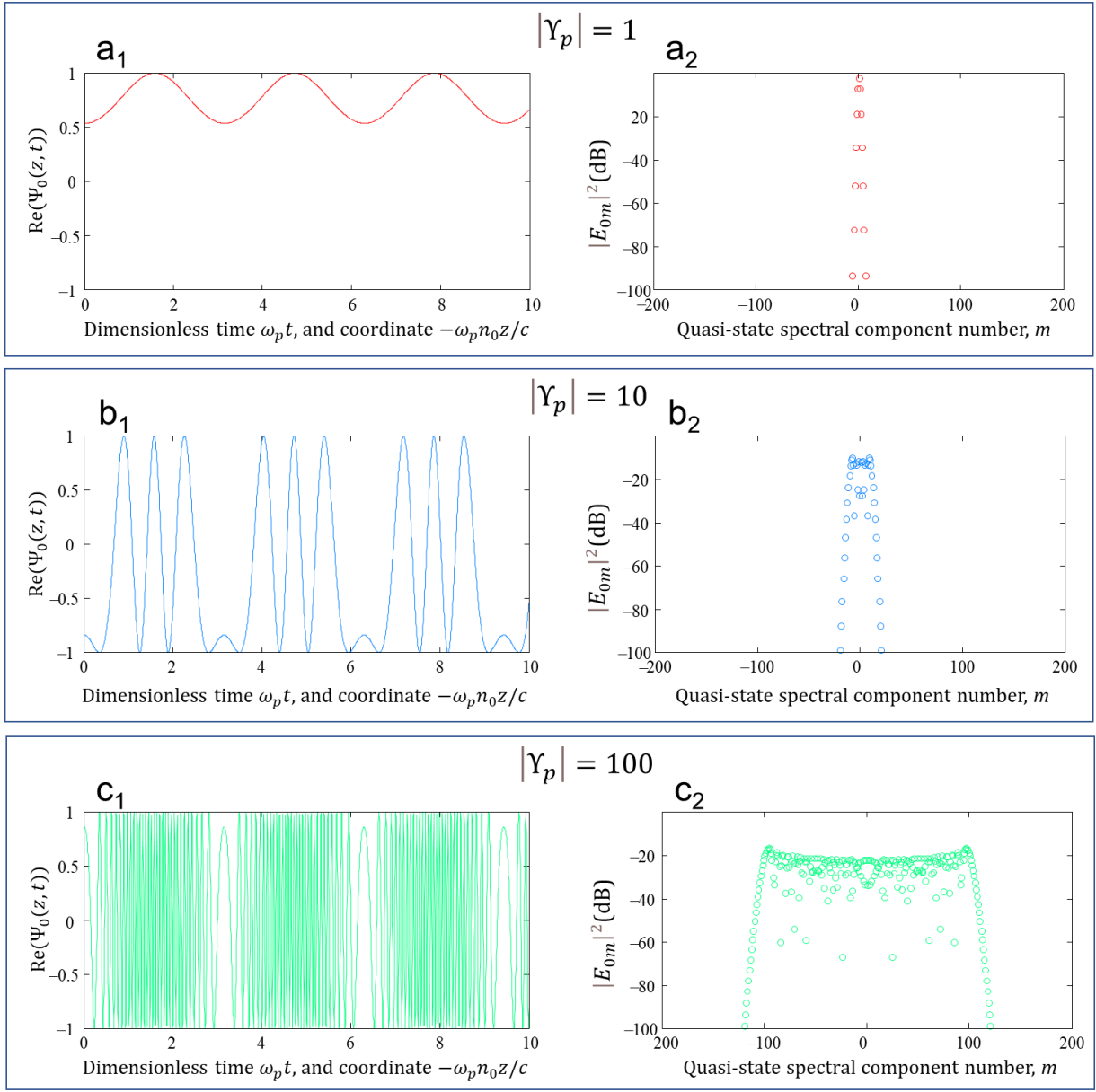

**Fig. D1.** Variation of the real part of the quasi-states (a$_1$, b$_1$, and c$_1$) and comb line amplitudes (a$_2$, b$_2$, and c$_2$) of a standing along RTM for different values of the modulation indices ($|\Upsilon_p| = 1$, 10, and 100).

Near the condition of parametric resonance, $\omega_0 T = 2\pi N$, $N = 1, 2, 3 \ldots$, and for the adiabatically slow modulation, $\omega_0 T \ll 1$, the modulation index magnitude $|\Upsilon_p|$ is large. Then, the characteristic frequency of oscillations of $\Psi_q(z,t)$ increases and becomes much greater than $\omega_p$. This results in growth of the power and bandwidth of the equidistant comb spectrum of $E_q(z,t)$ determined by Eqs. (2.11) and (2.13). To illustrate this, Figs. D1(a$_1$), (b$_1$), and (c$_1$) show the plots of $\text{Re}(\Psi_0(z,t))$ for $|\Upsilon_p| = 1, 10,$ and $100$. Respectively, Fig. D1(a$_2$), (b$_2$) and (c$_2$) show the power spectra of $|E_{0m}|$ for the same parameters. It is seen that, provided the modulation index and, consequently, the quasi-state frequency of oscillations grow significantly, the comb spectrum becomes wide and nicely flattened.

Generally, the field inside a weakly-coupled driven microresonator can be presented as a linear combination of its quasi-states. For sufficiently weak coupling and losses, a coherent external source can excite a single quasi-state provided that the eigenstates are not degenerated, i.e., under the condition that there are no identical frequencies among frequencies $\omega_0 + \frac{2\pi q}{T} - \omega_p m$ of harmonics in Eq. (D9). At modulation resonances and adiabatic condition, $\omega_p T \to 2\pi N$, $N = 0, 1, 2, \ldots$, the eigenstates become degenerated while the magnitude of modulation index $|\Upsilon_p| \to \infty$. Then, the field excited inside the microresonator by the external coherent source strongly depends on the spatial distribution of the source.

Special situation occurs in the case of a uniform SDM when $\Delta n_{p1}(x) = \Delta n_{p1}^0 = const$ along the whole resonator length. Then we find from Eq. (D8) and Eq. (6) that the quasi-state modulation index is equal to $\Upsilon_p = -\frac{2i\omega_0 \Delta n_{p1}^0}{n_0 \omega_p}$. From this equation, we find that though $\Upsilon_p$ tends to infinity in the adiabatic limit $\omega_p T \to 0$, it does not have singularities at modulation resonances.

**Appendix E. Asymptotic calculation of the comb amplitude for the RTM coupled to a waveguide**

Taking into account that for small coupling $\kappa \ll 1$ considered here we have $\tau = 1 - \frac{\kappa^2}{2}$, the comb line amplitude in Eq. (4) for $m \geq 1$ can be written down as

$$|E_m| = \kappa^2 \left| \sum_{n=0}^{\infty} J_m\left(\sigma_{n+1} |\Omega_{p1}|\right) \exp\left[(n+1)\left(-\frac{im}{2}\omega_p T + i\omega_{in} T - \alpha + i\Omega_{p0}\right)\right] \right|,$$

$$\alpha = \frac{\kappa^2}{2} + \frac{2\eta \omega L}{c}.$$
(E1)

We are interested in situations when the amplitude of $|E_m|$ with large numbers $m \gg 1$ is resonantly enhanced. This can be achieved only if many terms with numbers $n \gg 1$ contribute to the sum in $|E_m|$. We suggest the interval where these terms have close phases (and, thus, contribute constructively) is within the interval $0 < m < \sigma_{n+1}|\Omega_{p1}|$ since Bessel functions in Eq. (E1) vanish with growing $m$ for $m > \sigma_{n+1}$. We assume that $\omega_p T$ in Eq. (E1) is resonant or adiabatically small, i.e., $\omega_p T = 2\pi N + \Delta\omega_p T, n|\Delta\omega_p T| \ll 1, N = 0, 1, 2, \ldots$. Then

$$\sigma_{n+1} \cong (-1)^{Nn}(n+1).$$
(E2)

Since the number of terms contributing to the sum in Eq. (E1) is $\sim \alpha^{-1}$, the condition of validity of Eq. (E2) is $|\Delta\omega_p T| \ll \alpha$. First, we exclude the adiabatic case, i.e., assume $N > 0$. Substitution of $\omega_p T = 2\pi N + \Delta\omega_p T$, Eq. (E2), and the asymptotic expansion for the Bessel function with large order and argument,

$$J_m\left(\sigma_{n+1}|\Omega_{p1}|\right) \cong \left(\frac{2}{\pi}\right)^{1/2} \left(\sigma_{n+1}^2 |\Omega_{p1}|^2 - m^2\right)^{-1/4} e^{i\pi Nmn} \cos\left(\sqrt{\sigma_{n+1}^2 |\Omega_{p1}|^2 - m^2} - m \cdot \text{acos}\left(\frac{m}{x}\right) - \frac{\pi}{4}\right)$$
(E3)

into Eq. (E1) yields:

$$|E_m| \cong (2\pi)^{-1/2} \kappa^2 \left| \sum_{\pm} \exp\left(\mp \frac{i\pi}{4}\right) \sum_{n=1}^{\infty} \left(n^2 |\Omega_{p1}|^2 - m^2\right)^{-1/4} \exp\left(i\varphi^{\pm}(n)\right) \right|,$$

$$\varphi^{\pm}(n) = \pm\sqrt{n^2 |\Omega_{p1}|^2 - m^2} \mp m \cdot \text{acos}\left(\frac{m}{n|\Omega_{p1}|}\right) + n\left[(\omega_{in} - \omega_s)T - \frac{1}{2}m\Delta\omega_p T + i\alpha^+\right].$$
(E4)

Here we introduced the resonant frequencies $\omega = \omega_s$ satisfying the quantization condition

$$\omega_s T + \Omega_{p0} = 2\pi s, \quad s \gg 1, \text{ integer.} \tag{E5}$$

This equation is identical to Eq. (10). The resonant condition takes place when a sufficiently large number of terms in Eq. (E4) behave smoothly as a function of $n$, have close phases, and, thus, contribute constructively to $I_m$. This happens near the stationary point $n = n_m^\pm$ determined by the equation

$$\frac{d\varphi^\pm(n)}{dn} = \pm\frac{1}{n}\sqrt{n^2|\Omega_{p1}|^2 - m^2} + (\omega_{in} - \omega_s)T - \frac{1}{2}m\Delta\omega_p T + i\alpha = 0, \tag{E6}$$

which formally results in equal complex-valued saddle points

$$n_m^\pm = \frac{m}{|\Omega_{p1}|\sqrt{1-\Lambda_m^2}}, \quad \Lambda_m = \frac{(\omega_{in} - \omega_s)T - \frac{1}{2}m\Delta\omega_p T + i\alpha^+}{|\Omega_{p1}|}. \tag{E7}$$

Here, the required positive value of the real part of $n_m^\pm$ can be ensured by selecting the appropriate signs + or − in Eq. (E6) (and, thus, terms corresponding to sign + or – in the sum $\Sigma_\pm$ in Eq. (E4)) depending on whether the deviation of $\omega$ from $\omega_s$ is negative or positive, respectfully. Next, choosing the appropriate sign of the square roots, we find:

$$\varphi^\pm(n_m^\pm) = m \cdot \text{acos}\left(\sqrt{1-\Lambda_m^2}\right) = -i \cdot m \ln\left(\sqrt{1-\Lambda_m^2} + i\Lambda_m\right),$$

$$\left|\frac{d^2\varphi^\pm(n)}{dn^2}\right|_{n=n_m^\pm} = \left|\frac{m^2}{n^2\sqrt{n^2|\Omega_{p1}|^2 - m^2}}\right|_{n=n_m^\pm} = \frac{|\Omega_{p1}|\left|1-\Lambda_m^2\right|^{3/2}}{m|\Lambda_m|}. \tag{E8}$$

Using these equations, we calculate the sum in Eq. (E4) by the saddle point method and finally get:

$$|E_m| \cong \kappa^2 \left|\Omega_{p1}\left(1-\Lambda_m^2\right)\right|^{-1/2} \left|\Lambda_m - i\sqrt{1-\Lambda_m^2}\right|^{|m|}. \tag{E9}$$

For the weak modulation index, $|\Omega_{p1}| \ll 1$, Eq. (E9) can be found directly from Eq. (9) of the main text by replacing $\exp(i|\Omega_{p1}|\cos(...))$ by $1 + i|\Omega_{p1}|\cos(...)$ and expanding the denominator in Eq. (9) into Fourier series [49]:

$$\frac{1}{a+b\cos(x)} = \frac{1}{\sqrt{a^2-b^2}} \sum_{n=-\infty}^{\infty} \left(\sqrt{\left(\frac{a}{b}\right)^2-1} - \frac{a}{b}\right)^{|n|} \exp(inx). \tag{E10}$$

## Appendix F. Solution of the first order partial differential equation describing semiclassical propagation of light along an SBM

Our analysis of the quasi-states and frequency comb generation by SBMs is based on the solutions $E^\rightleftarrows(z,t)$ and $\Psi^\rightleftarrows(z,t)$ of Eq. (15) where indices → and ← correspond to the propagation of light along the axis $z$ into the positive and negative directions along $z$, respectively:

$$E^{\rightleftarrows}(z,\omega,t) = \frac{1}{\sqrt{\beta(z)}} \exp[-i\omega t \pm iS(z,\omega)]\Psi^{\pm}(z,\omega,t),$$

$$\Psi^{\rightleftarrows}(z,\omega,t) = \frac{1}{\sqrt{\beta(z)}} \exp(\pm i\Delta S^{\pm}(z,\omega,t))\Phi^{\rightleftarrows}(t \mp \tau(z,\omega)). \tag{F1}$$

Here $\omega$ is an arbitrary constant, $\Phi^{\rightleftarrows}(t)$ are arbitrary relatively slow function of time $t$, and propagation constant $\beta(z)$, propagation time $\tau(z)$, action $S(z)$, and action increments $\Delta S^{\rightleftarrows}(z,t)$, are determined by equations [42]:

$$\beta(z,\omega) = \sqrt{\frac{2}{\chi}(\Delta\omega - \Delta\omega_{cut}(z))}, \quad \Delta\omega = \omega - \omega_0, \quad \Delta\omega_{cut}(z) = \omega_{cut}(z) - \omega_0, \quad \chi = \frac{c^2}{n_0^2 \omega_0},$$

$$\tau(z,\omega) = \frac{1}{\chi}\int_0^z \frac{dz}{\beta(z,\omega)}, \quad S(z,\omega) = \int_0^z \beta(z,\omega)dz, \tag{F2}$$

$$\Delta S^{\rightleftarrows}(z,\omega,t) = \frac{1}{\chi}\int_0^z \Delta\omega_p\left(z', t \mp \tau(z) \pm \tau(z')\right)\frac{dz'}{\beta(z',\omega)} + i\frac{\gamma}{\chi}\int_0^z \frac{dz'}{\beta(z',\omega)}.$$

The general solution defined by Eqs. (F1) and (F2) is valid for

$$|\Delta S^{\rightleftarrows}(z,\omega,t)| \ll |S(z,\omega)|. \tag{F3}$$

**Appendix G. Uncoupled SBM with semiclassical boundary conditions at the edges**

We look for the quasi-states of the uncoupled SBM numerated by quantum number $q$ in the form of a linear combination of fields $E^{\rightarrow}(x,t)$ and $E^{\leftarrow}(x,t)$ defined by Eqs. (F1) and (F2):

$$E_q(z,\omega,t) = E^{\rightarrow}(z,\omega,t) + E^{\leftarrow}(z,\omega,t) \tag{G1}$$

Assuming that the CF is smooth along the whole SBM length including edges, we determine the semiclassical boundary conditions for $E^{\rightleftarrows}(z,t)$ at the SBM edges corresponding to the semiclassical turning points at $z = 0$ and $z = L$: $E^{\leftarrow}(0,\omega,t) = E^{\rightarrow}(0,\omega,t)\exp\left(\frac{i\pi}{2}\right)$ and $E^{\leftarrow}(L,\omega,t) = E^{\rightarrow}(L,\omega,t)\exp\left(-\frac{i\pi}{2}\right)$ [7]. Then we find from Eqs. (F1) and (F2):

$$\Phi^{\leftarrow}(t) = \Phi^{\rightarrow}(t)\exp\left(\frac{i\pi}{2}\right),$$

$$\Phi^{\rightarrow}\left(t + \frac{T}{2}\right)\exp[-2iS(L,\omega) - i\Delta S(L,\omega,t) + \gamma T] = \Phi^{\leftarrow}\left(t - \frac{T}{2}\right)\exp\left(-\frac{i\pi}{2}\right), \tag{G2}$$

$$\Delta S(L,\omega,t) = \Delta S^{\rightarrow}(L,\omega,t) + \Delta S^{\leftarrow}(L,\omega,t), \quad T = 2\tau(L,\omega).$$

Here $T$ is the full roundtrip propagation time. Eq. (G2) leads to the functional equation:

$$\Phi^{\rightarrow}\left(t + \frac{T}{2}\right) = -\Phi^{\rightarrow}\left(t - \frac{T}{2}\right)\exp[2iS(L,\omega) + i\Delta S(L,\omega,t) - \gamma T]. \tag{G3}$$

For the harmonic modulation determined by Eq. (14),

$$\Delta\omega_p(z,t) = -i\gamma + \Delta\omega_{p0}(z) + \Delta\omega_{p1}(z)\cos(\omega_p t), \quad |\Delta\omega(z,t)| \ll |\Delta\omega_{cut}^0(z)|,$$

we have

$$\Delta S(L,\omega,t) = \tilde{\Omega}_{p0} + \tilde{\Omega}_{p1}\cos(\omega_p t) + i\gamma T$$

$$\tilde{\Omega}_{p0} = \frac{2}{\chi}\int_0^L \Delta\omega_{p0}(z)\frac{dz}{\beta(z,\omega)}, \quad \tilde{\Omega}_{p1} = \frac{2}{\chi}\int_0^L \Delta\omega_{p1}(z)\cos\left(\omega_p\left(\frac{T}{2}-\tau(z)\right)\right)\frac{dz}{\beta(z,\omega)}. \tag{G4}$$

It is convenient to fix $\omega_0$ in the vicinity of CF $\omega_{cut}(z,t)$ by the quantization rule [7]:

$$2S(L,\omega_0) + \Omega_{p0} = 2\pi\left(n_0 + \frac{1}{2}\right), \quad n_0 \gg 1, \text{ integer}. \tag{G5}$$

Here the integer $n_0$ is chosen to minimize the magnitude of $\Delta\omega_{cut}^0(z)$. In the absence of modulation, $\tilde{\Omega}_{p0} = \tilde{\Omega}_{p1} = 0$, and material and scattering losses, $\gamma = 0$, this rule determines the actual axial eigenfrequencies of the SBM. For small deviation of frequency $\omega$ from $\omega_0$, we have

$$S(z,\omega) = S(z,\omega_0) + \Delta\omega\tau(z,\omega_0), \quad \Delta\omega = \omega - \omega_0. \tag{G6}$$

Combining Eqs. (G3), (G5), and (G6) we have

$$\Phi^{\rightarrow}\left(t+\frac{T}{2}\right) = \Phi^{\rightarrow}\left(t-\frac{T}{2}\right)\exp\left(i\Delta\omega T - \gamma T + i\tilde{\Omega}_{p1}\cos(\omega_p t)\right). \tag{G7}$$

To solve Eq. (G7) we introduce function $f^{\rightarrow}(t)$ so that

$$\Phi^{\rightarrow}(t) = \exp\left(if^{\rightarrow}(t)\right) \tag{G8}$$

and rewrite Eq. (G7) as

$$f^{\rightarrow}\left(t+\frac{T}{2}\right) - f^{\rightarrow}\left(t-\frac{T}{2}\right) = (\Delta\omega + i\gamma)T + \tilde{\Omega}_{p1}\cos(\omega_p t) + 2\pi q, \quad q = 0, \pm 1, \pm 2, \ldots. \tag{G9}$$

Solution of this equation is

$$f^{\rightarrow}(t) = f_0\sin(\omega_p t) + \xi t, \quad f_0 = \frac{\tilde{\Omega}_{p1}}{2\sin\left(\frac{\omega_p T}{2}\right)}, \quad \xi = \Delta\omega + i\gamma + \frac{2\pi q}{T}. \tag{G10}$$

It follows from Eq. (G10) that

$$\Phi^{\rightarrow}(t) = \exp\left(if^{\rightarrow}(t)\right) = \exp\left(f_0\sin(\omega_p t) + \xi t\right), \quad \Phi^{\leftarrow}(t) = \Phi^{\rightarrow}(t)\exp\left(\frac{i\pi}{2}\right). \tag{G11}$$

Substitution of these expressions for $\Phi^{\rightleftarrows}(t)$ into Eq. (G1) leads to the exclusion of undefined frequency shift $\Delta\omega = \omega - \omega_0$ and yields the expression for normalized quasi-eigenfrequencies $\omega_q$ and quasi-states $E_q(z,t)$ of a free SBM:

$$\omega_q = \omega_0 + \frac{2\pi q}{T}, \quad q = 0, \pm 1, \pm 2, \ldots,$$

$$E_q(z,t) = \frac{C}{\sqrt{\beta_0(z)}} \exp\left[-i\left(\omega_0 + i\gamma + \frac{2\pi q}{T}\right)t + i\frac{\tilde{\Omega}_{p1}}{2\sin\left(\frac{\omega_p T}{2}\right)} \sin\left(\omega_p\left(t - \tau_0(z)\right)\right)\right] \times \quad\quad (G12)$$

$$\cos\left(\frac{\pi}{4} + S_0(z) + \frac{2q\pi}{T}\tau_0(z) + \Delta S_0(z,t)\right),$$

where

$$C = \sqrt{2}\left(\int_0^L \frac{dz}{\beta_0(z)}\right)^{-1/2}, \quad S_0(z) = \int_0^z \beta_0(z)dz, \quad \beta_0(z) = \sqrt{-\frac{2}{\chi}\Delta\omega_{cut}(z)}, \quad \tau_0(z) = \frac{1}{\chi}\int_0^z \frac{dz}{\beta_0(z)},$$

$$\Delta S_0(z,t) = \frac{2}{\chi}\int_0^z \Delta\omega_{p0}(z)\frac{dz}{\beta_0(z)} + \frac{2}{\chi}\cos(\omega_p t)\int_0^z \Delta\omega_{p1}(z)\cos\left(\omega_p\left(\frac{T}{2} - \tau_0(z)\right)\right)\frac{dz}{\beta_0(z)}, \quad\quad (G13)$$

$$\Delta S_0(L,t) = \tilde{\Omega}_{p0} + \tilde{\Omega}_{p1}\cos(\omega_p t),$$

$$\tilde{\Omega}_{p0} = \frac{2}{\chi}\int_0^L \Delta\omega_{p0}(z)\frac{dz}{\beta_0(z)}, \quad \tilde{\Omega}_{p1} = \frac{2}{\chi}\int_0^L \Delta\omega_{p1}(z)\cos\left(\omega_p\left(\frac{T}{2} - \tau_0(z)\right)\right)\frac{dz}{\beta_0(z)}.$$

**Appendix H. Rectangular SBM**

We consider an SBR with a rectangular CF which is uniform in the region between its edges $0 < z < L$,

$$\Delta\omega_{cut}(z) \equiv \Delta\omega_{cut}^0, \quad\quad (H1)$$

and has abrupt and sufficiently large negative breaks at edges $z = 0$ and $z = L$ leading to the boundary conditions $E_q(0, t) = 0$ and $E_q(L, t) = 0$. We look for the quasi-states in the form given by Eq. (G1) so that these boundary conditions are equivalent to conditions $E^{\leftarrow}(0, \omega, t) = -E^{\rightarrow}(0, \omega, t)$ and $E^{\leftarrow}(L, \omega, t) = -E^{\rightarrow}(L, \omega, t)$. The expressions given by Eqs. (16), (18), and (22) for the propagation constant, circulation time, and modulation parameters are now simplified to

$$\beta(z,\omega_0) = \beta_0 = \sqrt{\frac{2}{\chi}\Delta\omega_{cut}^0} = \frac{n_0}{c}\sqrt{2\omega_0\Delta\omega_{cut}^0}, \quad \tau(z) = \frac{z}{\chi\beta_0}, \quad T = \tau(2L,\omega_0) = \frac{n_0 L}{c}\sqrt{\frac{2\omega}{\Delta\omega_{cut}^0}},$$

$$\tilde{\Omega}_{p0} = \frac{2}{\chi\beta_0}\int_0^L \Delta\omega_{p0}(z)dz, \quad \tilde{\Omega}_{p1} = \frac{2}{\chi\beta_0}\int_0^L \Delta\omega_{p1}(z)\cos\left(i\omega_p\left(\frac{T}{2} - \frac{z}{\chi\beta_0}\right)\right)dz, \quad \chi = \frac{c^2}{n_0^2 \omega_0}. \quad\quad (H2)$$

We fix $\omega_0$ in the vicinity of CF $\omega_{cut}(z, t)$ by the quantization rule (compare with (G5)):

$$2S(L,\omega_0) + \Omega_{p0} = 2\pi n_0, \quad n_0 \gg 1, \text{ integer.} \quad\quad (H3)$$

Then, the equations for functions $\Phi^{\rightarrow}(t)$ and $\Phi^{\leftarrow}(t)$ become:

$$\Phi^{\leftarrow}(t) = -\Phi^{\rightarrow}(t),$$

$$\Phi^{\leftarrow}\left(t + \frac{T}{2}\right)\exp\left[-2iS(L,\omega_0) - i\Delta S(L,\omega_0,t)\right] = -\Phi^{\rightarrow}\left(t - \frac{T}{2}\right) \quad\quad (H4)$$

Eqs. (H4), (H3), (G4), and (G6) lead to Eq. (G7) for $\Phi^{\rightarrow}(t)$ which solution determines $\Phi^{\rightarrow}(t)$ and $\Phi^{\leftarrow}(t)$ by Eqs. (G8), (G10), and (G11). Finally, the expression for the quasi-states of the rectangular SBM is

$$E_q(z,t) = \sqrt{\frac{2}{L}} \exp\left[-i\left(\omega_0 + i\gamma + \frac{2\pi q}{T}\right)t + i\frac{\tilde{\Omega}_{p1}}{2\sin\left(\frac{\omega_p T}{2}\right)} \sin\left(\omega_p\left(t - \frac{z}{\chi\beta_0}\right)\right)\right] \cos\left(\beta_0 z + \frac{2q\pi z}{\chi\beta_0 T} + \Delta S_0(z,t)\right), \quad \text{(H5)}$$

where

$$\Delta S_0(z,t) = \frac{2}{\chi\beta_0} \int_0^z \Delta\omega_{p0}(z)dz + \frac{2}{\chi\beta_0} \cos(\omega_p t) \int_0^z \Delta\omega_{p1}(z) \cos\left(\omega_p\left(\frac{T}{2} - \frac{z}{\chi\beta_0}\right)\right) dz. \quad \text{(H6)}$$

## Appendix I. An SBM coupled to the input-output waveguide

We rewrite Eq. (19) as

$$E_{out}(t) = \tau E_{in}(t) + \kappa E^{\leftarrow}(0,t) \quad \text{(I1)}$$

$$E^{\rightarrow}(0,t) = -\kappa E_{in}(t) + \tau E^{\leftarrow}(0,t) \quad \text{(I2)}$$

Substitution of the input fields

$$E_{in}(t) = \exp(-i\omega t), \quad \text{(I3)}$$

$$E_{out}(t) = \exp(-i\omega t) E_{out}^0(t) \quad \text{(I4)}$$

and

$$E^{\rightleftarrows}(0,t) = \exp(-i\omega t)\Phi^{\pm}(t) \quad \text{(I5)}$$

into Eqs. (I1) and (I2), yields the functional equations for $\Phi^{\pm}(t)$:

$$\begin{aligned} E_{out}(t) &= \tau + \kappa\Phi^{-}(t) \\ \Phi^{+}(t) &= -\kappa + \tau\Phi^{-}(t) \\ \Phi^{-}(t)\exp &= \Phi^{+}(t-T)\tilde{A}(t), \end{aligned} \quad \text{(I6)}$$

where

$$\tilde{A}(t) = \exp\left(2i\int_0^L \beta(z)dz + i\tilde{\Omega}_{p0} + i\tilde{\Omega}_{p1}\cos\left(\omega_p\left(t - \frac{T}{2}\right)\right) - \frac{i\pi}{2}\right). \quad \text{(I7)}$$

The functional equation for $\Phi^{+}(t)$ following from Eqs. (I6) is:

$$\Phi^{+}\left(t + \frac{T}{2}\right) = \tau\tilde{A}(t)\Phi^{+}\left(t - \frac{T}{2}\right) - \kappa. \quad \text{(I8)}$$

Introducing $\tilde{\Phi}(t) = \Phi\left(t + \frac{T}{2}\right)$, we reduce Eq. (I8) to the one identical to Eq. (C.3)

$$\tilde{\Phi}^{+}(t) = \tau \tilde{\Phi}^{+}(t-T) \exp\left[2iS(L) + i\Delta S\left(L, t-\frac{T}{2}\right) - \frac{i\pi}{2}\right] - \kappa. \tag{I9}$$

and find the solution of Eq. (I8) similar to that given by Eq. (C.4):

$$\Phi^{+}(t) = -\kappa\left(1 + \sum_{n=1}^{\infty} \tau^n \prod_{m=0}^{n-1} \tilde{A}(t-mT)\right),$$

$$\tilde{A}(t) = \exp\left[2iS(L) + i\Delta S\left(L, t-\frac{T}{2}\right) - \frac{i\pi}{2}\right]. \tag{I10}$$

After substitution of Eqs. (I1)-(I5) into Eq. (I10), we find

$$E_{out}(t) = E_{in}(t)\tau\left\{1 - \kappa^2 \sum_{n=0}^{\infty} \tau^n \exp\left[(n+1)\left(i\omega T - \frac{2\eta\omega L}{c} + i\Omega_{p0}\right)\right.\right.$$
$$\left.\left. + i\sigma_{n+1}\cos\left(\omega_p t - \frac{n+2}{2}\omega_p T + \arg(\Omega_{p1})\right)\right]\right\}. \tag{I11}$$

Expanding $\exp(i\sigma_{n+1}\cos(\ldots))$ in this equation into series using Jacobi-Anger formula leads to Eq. (21) of the main text.

### Appendix J. Optimization of the power consumption: formulation

First, we describe the method of optimization of power consumption for the RTM. We consider the resonant parametric modulation when $\omega_p T = 2\omega_p n_0 L/c = 2\pi N$, $N = 1, 2, \ldots$ having the modulation index ( see Eq. (6) of the main text)

$$\Omega_{p1} = \frac{\omega_0}{c} \int_0^{2L} dz \Delta n_{p1}(z) \exp\left(\frac{i\pi N}{L} z\right), \tag{J1}$$

Then, for the Pockels modulation the power consumption can be presented as

$$P_{Pock} = W_{Pock} \int_0^{2L} dz \Delta n_{p1}^2(z). \tag{J2}$$

with constant proportionality coefficient $W_{Pock}$.

Presenting the modulation amplitude of refractive index as $n_{p1}(z) = n_{p1}^0 f(z)$, expressing $n_{p1}^0$ through $P_{Pock}$ from Eq. (J2) as $n_{p1} = P_{Pock}^{1/2}\left(W_{Pock}\int_0^{2L} dz f^2(z)\right)^{-1/2}$ and substituting it into (J1), we have:

$$|\Omega_{p1}| = \frac{\omega_0}{c}\sqrt{\frac{P_{Pock}}{W_{Pock}}} \frac{\left|\int_0^{2L} dz \Delta n_{p1}(z)\exp\left(\frac{i\pi N}{L} z\right)\right|}{\sqrt{\int_0^{2L} dz \Delta n_{p1}^2(z)}}. \tag{J3}$$

Thus, to maximise the modulation index $|\Omega_{p1}|$ for the fixed power consumption $P_{Pock}$, we have to maximise the last fraction in this expression, which, within a constant factor, coincides with the expression for $\Omega_{p1}^{Pock}$ in Eq. (42).

Similar, for the Kerr modulation we have

$$P_{Kerr} = W_{Kerr} \int_0^{2L} dz \left| \Delta n_{p1}(z) \right|. \tag{J4}$$

Similar to the Pockels modulation, after the substitution $n_{p1}(z) = n_{p1}^0 f(z)$, expressing $n_{p1}^0$ through $P_{Kerr}$ from Eq. (J4) and substituting it into (J3), we have:

$$\left|\Omega_{p1}\right| = \frac{\omega_0}{c} \frac{P_{Kerr}}{W_{Kerr}} \frac{\left| \int_0^{2L} dz \Delta n_{p1}(z) \exp\left(\frac{i\pi N}{L} z\right) \right|}{\int_0^{2L} dz \left|\Delta n_{p1}(z)\right|}. \tag{J5}$$

For the fixed power consumption $P_{Kerr}$, we have to maximize the last fraction in the expression which, within a constant factor, coincides with the expression for $\Omega_{p1}^{Kerr}$ in Eq. (42).

For the SBM with arbitrary shape, the expression for the modulation index can be written down as (see Eqs. (22) and (26) of the main text)

$$\tilde{\Omega}_{p1} = \int_0^{\tilde{L}} \Delta n_{p1}(z) G(z) dz, \quad G(z) = \frac{2\omega_0}{\chi n_0 \beta(z,\omega_0)} \cos\left(\tilde{\omega}_p \left(\frac{\tilde{T}}{2} - \frac{1}{\chi} \int_0^z \frac{dz}{\beta(z,\omega_0)}\right)\right), \tag{J6}$$

and the optimization is performed similarly to the RTM using equations analogous to Eqs. (J2)-(J5). In particular, to find the optimum $\widetilde{\Delta n}_{p1}(z)$ in the case of the Pockels modulation, we have to maximize the last fraction in the expression (compare with Eq. (J3)):

$$\left|\tilde{\Omega}_{p1}\right| = \sqrt{\frac{P_{Pock}}{W_{Pock}}} \frac{\left|\int_0^{\tilde{L}} \Delta n_{p1}(z) G(z) dz\right|}{\sqrt{\int_0^L dz \Delta n_{p1}^2(z)}}. \tag{J7}$$

**Appendix K. Optimization of the power consumption for Pockels modulation: exact solution**

The optimized spatial profile of parametric modulation determined in Appendix J can be found analytically. For this purpose, starting with an SBM, we find the optimum $\widetilde{\Delta n}_{p1}(z)$ maximizing the absolute value of modulation index defined by Eq. (J7) using the Cauchy-Schwarz inequality:

$$\left(\int_0^{\tilde{L}} \Delta n_{p1}(z) G(z) dz\right)^2 \leq \int_0^L dz \Delta n_{p1}^2(z) \int_0^L dz G^2(z). \tag{K1}$$

It immediately follows from this inequality and Eq. (J7) that the maximum of $|\tilde{\Omega}_{p1}|$ is achieved for

$$\widetilde{\Delta n}_{p1}(z) \sim G(z). \tag{K2}$$

For the Pockels modulation of an RTM, the approach based on the inequality in Eq. (K1) fails since then the function similar to $G(z)$ in Eq. (J6) is proportional to the complex exponent (see Eq. (J1) for $\Omega_{p1}$) and the Cauchi-Schwarz inequality is reduced to a trivial relation. In this case, we solve this problem as follows. Assume that $\Delta n_{p1}(z)$ is the SDM maximizing $\left|\Omega_{p1}\right|$ defined by Eq. (J3) with fixed power $P$. Then we can always find a shift $z_t$ such that

$$s(z_t) = \int_0^{2L} dz \Delta n_{p1}(z) \sin\left(\frac{\pi N}{L}(z - z_t)\right) = 0. \tag{K3}$$

Obviously, $s(0) = -s(L/N)$. Therefore, since $s(z_t)$ is continuous, there exists $z_t$ such that $f(z_t) = 0$ (Bolzano's intermediate value theorem). For this $z_t$ we have

$$\Omega_{p1} = \frac{\omega_0}{c} \int_0^{2L} dz \Delta n_{p1}(z) \cos\left(\frac{\pi N}{L}(z - z_t)\right). \tag{K4}$$

Now, using the Cauchy-Schwarz inequality

$$\left(\int_0^{2L} dz \Delta n_{p1}(z) \cos\left(\frac{\pi N}{L}(z + z_t)\right)\right)^2 \leq \int_0^{2L} dz \left(\Delta n_{p1}(z)\right)^2 \int_0^{2L} dz \cos^2\left(\frac{\pi N}{L}(z + z_t)\right), \tag{K5}$$

we conclude that the SDM which maximizes $|\Omega_{p1}|$ in Eq. (J3) is $\Delta n_{p1}(z) = \Delta n_{p1}^{(opt)}(z) = \Delta n_{p1}^0 \cos(\pi N(z + z_t)/L)$. However, for an arbitrary shift $z_0$, we have

$$\left|\int_0^{2L} dz \Delta n_{p1}^{(opt)}(z + z_0) \exp\left(\frac{i\pi N}{L}z\right)\right| = \left|\int_0^{2L} dz \Delta n_{p1}^{(opt)}(z) \exp\left(\frac{i\pi N}{L}z\right)\right| = \frac{1}{2} \Delta n_{p1}^0. \tag{K6}$$

Therefore, the general solution maximizing $|\Omega_{p1}|$ for the given power consumption $P$ is

$$\Delta n_{p1}(z) \sim \cos\left(\frac{\pi N}{L}(z - z_0)\right). \tag{K7}$$

with arbitrary shift $z_0$.

**Appendix L. An SBM coupled to the input-output waveguide positioned away from the SBM edges**

We consider an SBM coupled to an input-output waveguide arbitrarily positioned along the SBM axis as illustrated in Fig. L1. Solutions of Eq. (15) in the left- and right-hand sides of the SBM can be presented in the form defined by Eqs. (F1) and (F2). After application of the boundary condition at the turning point $z = 0$ the waves propagating towards positive ($\rightarrow$) and negative ($\leftarrow$) directions along the SBM at its left-hand side, $z < L/2$, are found in the form:

$$E_{left}^{\rightleftarrows}(z,t) = H^{\pm}(z,t)\Phi_{left}\left(t \mp \tilde{t}_c(z)\right), \quad E_{right}^{\rightleftarrows}(z,t) = H^{\pm}(z,t)\Phi_{right}\left(t \mp \tilde{t}_c(z)\right),$$

$$H^{\pm}(z,t) = \sqrt{\frac{\beta(z_c)}{\beta(z)}} \exp\left(-i\omega t \pm i\int_{z_c}^{z} \beta(z)dz + i\frac{1}{\chi}\int_{z_c}^{z} \Delta\omega_p\left(z', t \mp \tilde{t}_c(z) \pm \tilde{t}_c(z')\right)\frac{dz'}{\beta(z')} \mp \gamma\tilde{t}_c(z)\right), \tag{L1}$$

$$\tilde{t}_c(z) = \frac{1}{\chi}\left|\int_{z_c}^{z}\frac{dz'}{\beta(z')}\right|.$$

Here $\Phi_{left}\left(t \mp \tilde{t}_c(z)\right)$ and $\Phi_{right}\left(t \mp \tilde{t}_c(z)\right)$ are arbitrary functions. It follows from Eq. (L1) that at the coupling point

$$E_{left}^{\rightleftarrows}(z_c,t) = \exp(-i\omega t)\Phi_{left}^{\pm}(t), \quad E_{right}^{\rightleftarrows}(z_c,t) = \exp(-i\omega t)\Phi_{right}^{\pm}(t). \tag{L2}$$

The semiclassical condition of vanishing of the field to the left of the turning point $z = 0$ and to the right from the turning point $z = L$ yields [7]

$$\vec{E}_{left}(0^{\swarrow},t) = \overleftarrow{E}_{left}(0^{\swarrow},t)\exp\left(-\frac{i\pi}{2}\right), \quad \vec{E}_{right}(L^{\nwarrow},t) = \overleftarrow{E}_{right}(L^{\nwarrow},t)\exp\left(-\frac{i\pi}{2}\right). \tag{L3}$$

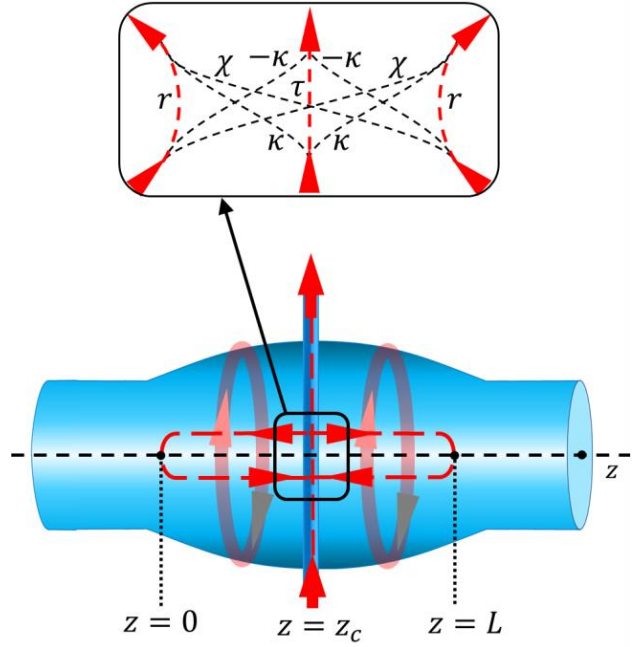

**Fig. L1.** An SBM coupled to a waveguide positioned at $z = z_c$ away from the WGM turning points $z = 0$ and $z = L$. Inset: parameters of S-matrix in Eq. (L7).

Here we denote the asymptotics of solutions near the right hand side of $z = 0$ and left hand side of $z = L$ by $0^{\swarrow}$ and $L^{\nwarrow}$, respectively. From Eqs. (L1) and (L3) we find

$$H^-\left(0^{\swarrow},t\right) = H^+\left(0^{\swarrow},t\right)\tilde{A}_{left}(t),$$

$$\tilde{A}_{left}(t) = \exp\left(\frac{i\pi}{2} + 2i\int_0^{z_c}\beta(z)dz + i\tilde{\Omega}_{p0}^{left} + i\tilde{\Omega}_{p1}^{left}\cos(\omega_p t) - 2\gamma T_{left}\right), \tag{L4}$$

$$\tilde{\Omega}_{p0}^{left} = \frac{2}{\chi}\int_0^{z_c}\frac{\Delta\omega_{p0}(z)}{\beta(z)}dz, \quad \tilde{\Omega}_{p1} = \frac{2}{\chi}\int_0^{z_c}\Delta\omega_{p1}(z)\cos\left(\omega_p\left(T_{left}-\tau(z)\right)\right)\frac{dz}{\beta(z)},$$

and

$$H^-\left(L^{\nwarrow},t\right) = H^+\left(L^{\nwarrow},t\right)\tilde{A}_{right}(t),$$

$$\tilde{A}_{right}(t) = \exp\left(-\frac{i\pi}{2} - 2i\int_{z_c}^{L}\beta(z)dz - i\tilde{\Omega}_{p0}^{right} - i\tilde{\Omega}_{p1}^{right}\cos(\omega_p t) - 2\gamma T_{right}\right), \tag{L5}$$

$$\tilde{\Omega}_{p0}^{right} = \frac{2}{\chi}\int_{z_c}^{L}\frac{\Delta\omega_{p0}(z)}{\beta(z)}dz, \quad \tilde{\Omega}_{p1} = \frac{2}{\chi}\int_{z_c}^{L}\Delta\omega_{p1}(z)\cos\left(\omega_p\left(T_{right}-\tau(z)\right)\right)\frac{dz}{\beta(z)}.$$

It follows from Eqs. (L1), (L4), and (L5) that

$$\tilde{A}_{left}(t)\Phi^+_{left}\left(t-T_{left}\right) = \Phi^-_{left}\left(t+T_{left}\right), \quad T_{left} = \tilde{t}_c(0),$$
$$\tilde{A}_{right}(t)\Phi^+_{right}\left(t-T_{right}\right) = \Phi^-_{right}\left(t+T_{right}\right), \quad T_{right} = \tilde{t}_c(L).$$
(L6)

As in Sections 2 and 3, we describe coupling between the input-output waveguide and SBM with the transfer matrix approach. The relations between the input wave, $E_{in}(t) = \exp(-i\omega t)$, the waves inside the SBM defined by Eq. (L1) at the waveguide position $z = z_c$, and the output wave $E_{out}(t) = E^{(0)}_{out}\exp(-i\omega t)$ are defined by the matrix equation:

$$\begin{pmatrix} E_{out}(t) \\ E^{\leftarrow}_{left}(z_c,t) \\ E^{\rightarrow}_{right}(z_c,t) \end{pmatrix} = S \begin{pmatrix} E_{in}(t) \\ E^{\rightarrow}_{left}(z_c,t) \\ E^{\leftarrow}_{right}(z_c,t) \end{pmatrix}, \quad S = \begin{pmatrix} \tau & \kappa & \kappa \\ -\kappa & \chi & r \\ -\kappa & r & \chi \end{pmatrix}$$
(L7)

The physical meaning of elements of S-matrix is illustrated in the inset of Fig. (L1). As in the previous cases, without loss of generality we assume that these elements are real [20]. It follows from the unitarity of S-matrix that

$$\tau^2 + 2\kappa^2 = 1,$$
$$\kappa^2 + \chi^2 + r^2 = 1,$$
$$r - \tau + \chi = 0,$$
$$\kappa^2 + 2\chi r = 0.$$
(L8)

These relations allow to express all elements of S-matrix through one of them. For example, for the assumed small waveguide-SBM coupling, $\kappa \ll 1$, we have:

$$r = -\frac{\kappa^2}{2}, \quad \tau = 1 - \kappa^2, \quad \chi = 1 - \frac{\kappa^2}{2}.$$
(L9)

Using Eq. (L2), we reduce Eq. (L7) to

$$E^{(0)}_{out}(t) = \tau + \kappa\left(\Phi^+_{left}(t) + \Phi^-_{right}(t)\right)$$
$$\Phi^-_{left}(t) = -\kappa + \chi\Phi^+_{left}(t) + r\Phi^-_{right}(t)$$
$$\Phi^+_{right}(t) = -\kappa + r\Phi^+_{left}(t) + \chi\Phi^-_{right}(t)$$
(L10)

The last two equations in Eq. (L10) together with two equations Eq. (L6) are four functional equations which determine the four unknown functions $\Phi^+_{left}(t), \Phi^-_{left}(t), \Phi^+_{right}(t)$, and $\Phi^-_{right}(t)$, while the first equation in Eq. (L10) determines the output field. The analytical solution of these equations has not been found here. However, these equations can be solved for the important case when the SBM is axially symmetric with respect to its center $z = L/2$, i.e., when $\Delta\omega^0_{cut}\left(\frac{L}{2} - z\right) = \Delta\omega^0_{cut}\left(\frac{L}{2} + z\right)$. We assume the excitation of SBM to be symmetric as well, so that $\Delta\omega_p\left(\frac{L}{2} - z, t\right) = \Delta\omega_p\left(\frac{L}{2} + z, t\right)$ and the input-output waveguide is positioned at the SBM center, $z_c = L/2$. Then, the field in the SBM can be ether axially symmetric or antisymmetric with respect to $z = L/2$. We are interested only in symmetric solutions since the antisymmetric solutions vanish at the SBM center and, therefore, do not couple to the waveguide. From Eq. (L2), the latter condition requires that $\Phi^{\pm}_{left}(t) = \Phi^{\mp}_{right}(t)$ and Eqs. (L6) and (L10) are simplified to:

$$E_{out}^{(0)}(t) = \tau + 2\kappa \Phi_{left}^{+}(t)$$

$$\tilde{A}_{left}\Phi_{left}^{+}\left(t - \frac{T}{4}\right) = -\kappa + (\chi + r)\Phi_{left}^{+}\left(t + \frac{T}{4}\right)$$

(L11)

The latter equation here is simply reduced to Eq. (B1) which is analytically solved in Appendix B.

### Appendix M. An RTM with an internal light source

The parametrically excited RTM with a light source is described by Eq. (A1). Then, the general solution of Eq. (A1) is determined by Eq. (A5) with arbitrary function $\Phi(t)$. For the harmonic modulation of refractive index determined by Eq. (2), the continuity condition $E(0, t) = E(2L, t)$ yields the functional equation for $\Phi(t)$:

$$\Phi(t) = A(t)\Phi(t - T) + B(t),$$

(M1)

where

$$A(t) = \exp\left[i\omega T + i\Omega_{p0} - \frac{\eta}{n_0}\omega T + i|\Omega_{p1}|\cos\left(\omega_p t + \arg(\Omega_{p1}) - \omega_p T\right)\right]$$

(M2)

and

$$B(t) = \frac{n_0 c}{2i\omega} A(t)\exp\left[-i\Delta\omega_{in}(t - T)\right]\overline{F_{in}}, \quad \overline{F_{in}} = \int_0^{2L} dx \exp\left(\frac{-i\omega_p n_0}{c}z\right)F_{in}(z).$$

(M3)

Function $\Phi(t)$ satisfying Eq. (M1) is found as (see Appendix B):

$$\Phi(t) = \frac{n_0 c}{2i\omega_0}\overline{F_{in}}\exp\left[-i\Delta\omega_{in}(t - T)\right]\left\{\exp\left[2i|\Omega_{p1}|\cos\left(\omega_p(t - T) + \arg(\Omega_{p1})\right)\right]\right.$$

$$\left. + \sum_{n=0}^{\infty}\exp\left(i(n+1)\Delta\omega_{in}T\right)\exp\left(2i|\Omega_{p1}|\sigma_{n+1}\cos\left(\omega_p t + \arg(\Omega_{p1}) - \frac{n+2}{2}\omega_p T\right)\right)\right\},$$

$$\sigma_n = \frac{\sin\left(\frac{n}{2}\omega_p T\right)}{\sin\left(\frac{1}{2}\omega_p T\right)}.$$

(M4)

Taking into account that $E(0, t) = \exp(-i\omega_0 t)\Phi(t)$, we find:

$$E(0, t) = \frac{n_0 c}{2i\omega_0}\overline{F_{in}}\exp\left(-i\omega_{in}t + i\Delta\omega_{in}T\right)\left\{\exp\left[2i|\Omega_{p1}|\cos\left(\omega_p(t - T) + \arg(\Omega_{p1})\right)\right]\right.$$

$$\left. + \sum_{n=0}^{\infty}\exp\left((n+1)\left(i\Delta\omega_{in}T - \frac{2\eta\omega L}{c}\right)\right)\exp\left(2i|\Omega_{p1}|\sigma_{n+1}\cos\left(\omega_p t + \arg(\Omega_{p1}) - \frac{n+2}{2}\omega_p T\right)\right)\right\}.$$

(M5)

We note the similarity of this expression for the field and Eq. (C8) for the output field for the RTM coupled to an input-output waveguide. This result justifies the approximation of a realistic input-output waveguide coupled to microresonator by an internal source used in several publications (see, e.g. [31, 46] and references therein).

**Data availability statement**

The data that support the findings of this study are available upon reasonable request from the authors.

**Acknowledgments**

The work on the paper was supported by the Engineering and Physical Sciences Research Council (EPSRC) (grants EP/P006183/1 and EP/W002868/1) and Leverhulme Trust (grant RPG-2022-014).